\documentclass[graybox]{svmult}

\usepackage{type1cm}        
\usepackage{makeidx}         
\usepackage{graphicx}        
\usepackage{multicol}        
\usepackage[bottom]{footmisc}

\usepackage{newtxtext}       %
\usepackage{newtxmath}       

\makeindex             

\usepackage{url}
\usepackage{xspace}
\usepackage{wrapfig}
\usepackage{amsmath}
\usepackage{mathrsfs}
\usepackage{pifont}
\usepackage{enumitem}
\usepackage{subcaption}
\usepackage{tikz}
\usetikzlibrary{calc}
\usetikzlibrary{positioning}
\usetikzlibrary{automata}
\usetikzlibrary{shapes.multipart}
\usetikzlibrary{shadows}
\usetikzlibrary{shapes}
\usetikzlibrary{decorations.pathreplacing}
\usetikzlibrary{backgrounds}
\usepackage{pgfplots}
\pgfplotsset{compat=newest}

\usepackage{listings}
\usepackage{algorithmic}

\newcommand{\tvcmt}[1]{{\color{magenta}{}}}
\newcommand{\pscmt}[1]{{\color{blue}{}}}
\newcommand{\vmcmt}[1]{{\color{red!70!black}{}}}
\newcommand{\changed}[1]{{#1}}

\newcommand{\kiki}[0]{$k$I$k$I\xspace}
\newcommand{\twols}{\textsc{2ls}\xspace}
\newcommand{\cbmc}{\textsc{Cbmc}\xspace}

\newcommand{\minisat}{\textsc{MiniSAT2}\xspace}

\newcommand{\verdict}[1]{\textsf{#1}}
\newcommand{\ok}[0]{\verdict{OK}}
\newcommand{\fail}[0]{\verdict{FAILURE}}
\newcommand{\unknown}[0]{\verdict{UNKNOWN}}

\newcommand{\true}{\mathit{true}}
\newcommand{\false}{\mathit{false}}

\newcommand{\x}{{x}} 

\newcommand{\vx}{{\vec{\x}}}
\newcommand{\vs}{{\vec{s}}}
\newcommand{\vxp}{{\vec{\x}'}}

\newcommand{\vdelta}{{\vec{\delta}}}
\newcommand{\vd}{{\vec{d}}}

\newcommand{\trans}{\mathit{Trans}}
\newcommand{\inv}{\mathit{Inv}}

\newcommand{\init}{\mathit{Init}}

\newcommand{\err}{\mathit{Err}}

\newcommand{\templ}{\mathcal{T}}

\newcommand{\ssavar}[3]{#1^{#2}_{#3}}
\newcommand{\ssa}[2]{\ssavar{#1}{}{#2}}
\newcommand{\ssaphi}[2]{\ssavar{#1}{\mathit{phi}}{#2}}
\newcommand{\ssalb}[2]{\ssavar{#1}{\mathit{lb}}{#2}}
\newcommand{\ssals}[2]{\ssavar{#1}{\mathit{ls}}{#2}}
\newcommand{\ssax}[1]{\ssa{x}{#1}}
\newcommand{\ssag}[1]{\ssa{g}{#1}}
\newcommand{\ssaxphi}[1]{\ssaphi{x}{#1}}
\newcommand{\ssaxlb}[1]{\ssalb{x}{#1}}
\newcommand{\ssaplb}[1]{\ssalb{p}{#1}}
\newcommand{\ssagls}[1]{\ssals{g}{#1}}
\newcommand{\ssagos}[1]{\ssavar{g}{os}{#1}}
\newcommand{\ssagfree}[1]{\ssavar{g}{\mathit{fr}}{#1}}
\newcommand{\ssagosij}[2]{\ssavar{g}{os}{#1,#2}}

\newcommand{\drfvar}[2]{dr\hspace*{-0.5mm}f(#1)_{#2}}

\newcommand{\dynobj}[2]{ao_{#1}^{#2}}
\newcommand{\dynobjset}[1]{AO_{#1}}

\newcommand{\nullptr}{\mathsf{null}}

\newcommand{\obj}{\mathit{Obj}}
\newcommand{\dobj}{\mathit{AO}}
\newcommand{\ndobj}{\mathit{NAO}}
\newcommand{\pdobj}{\mathit{PAO}}
\newcommand{\sdobj}{\mathit{SAO}}
\newcommand{\ptr}{\mathit{Ptr}}
\newcommand{\num}{\mathit{Num}}
\newcommand{\var}{\mathit{Var}}
\newcommand{\pvar}{\mathit{PVar}}
\newcommand{\nvar}{\mathit{NVar}}
\newcommand{\svar}{\mathit{SVar}}
\newcommand{\fld}{\mathit{Fld}}
\newcommand{\nfld}{\mathit{NFld}}
\newcommand{\pfld}{\mathit{PFld}}

\newcommand{\addr}{\mathit{Addr}}
\newcommand{\ptrlb}{\ptr^{lb}}

\newcommand{\freevar}{\mathit{fr}}

\renewcommand{\vec}[1]{{\boldsymbol{#1}}}

\newcommand{\vecv}[2]{\left(\begin{array}{c}{#1}\\{#2}\end{array}\right)}

\newcommand{\basetempl}{\widehat{\mathcal{T}}}
\newcommand{\rowexpr}{\mathit{e}}

\newcommand{\kinv}{\mathit{KInv}}

\newcommand{\adom}{\mathscr{A}}
\newcommand{\tdom}{\mathscr{T}}

\newcommand{\sprop}{\mathit{P}}
\newcommand{\strans}{\mathit{T}}
\newcommand{\sinv}{\mathit{I}}
\newcommand{\skinv}{\mathit{KInv}}

\newcommand{\serr}{\mathit{Err}}

\newcommand{\define}[1]{\emph{#1}}

\newcommand{\limplies}[0]{\Rightarrow}

\newcommand{\st}[0]{.}

\newcommand{\lte}[0]{\leqslant}

\newcommand{\bigland}[1]{\bigwedge_{#1}}

\newcommand{\result}[1]{{\small \color{gray} #1}}
\newcommand{\eqdef}{\equiv}

\newcommand{\rankparam}{\ell}

\newcommand{\rankparamvec}{L}

\newcommand{\rank}{\mathit{R}}
\newcommand{\rrank}{RR}
\newcommand{\lexrank}{LR}
\newcommand{\rranktempl}{\mathcal{RR}}
\newcommand{\lexranktempl}{\mathcal{LR}}

\makeatletter
\renewcommand{\verbatim@font}{\ttfamily\small}
\makeatother

\renewcommand{\descriptionlabel}[1]{\hspace{\labelsep}\textbf{#1}}

\begin{document}

\title*{2LS for Program Analysis}

\author{
  Daniel Kroening,
  Viktor Mal\'{\i}k,
  Peter Schrammel and
  Tom\'{a}\v{s} Vojnar
}

\institute{
  Daniel Kroening \at University of Oxford, United Kingdom and Diffblue Ltd,
    United Kingdom \email{kroening@cs.ox.ac.uk} \and
  Viktor Mal\'{\i}k \at Faculty of Information Technology, Brno University of
    Technology, Czech Republic and IT4Innovations Centre of Excellence, Czech Republic,
    \email{imalik@fit.vutbr.cz} \and
  Peter Schrammel \at University of Sussex, Brighton, United Kingdom and
    Diffblue Ltd, United Kingdom \email{p.schrammel@sussex.ac.uk} \and
  Tom\'{a}\v{s} Vojnar \at Faculty of Information Technology, Brno
    University of Technology, Czech Republic and IT4Innovations Centre of
    Excellence, Czech Republic, \email{vojnar@fit.vutbr.cz}
}

\maketitle

\section{Introduction}
\label{sec:introduction}

2LS ("tools") is a verification tool for C programs,
built upon the CPROVER framework. It allows one to verify user-specified
assertions, memory safety properties (e.g. buffer overflows),
numerical overflows, division by zero, memory leaks, and
termination properties.

The analysis is performed by translating the
verification task into a second-order logic formula over bitvector,
array, and floating-point arithmetic theories. The formula is solved
by a modular combination of algorithms involving unfolding and
template-based invariant synthesis with the help of incremental
SAT solving.

Advantages of 2LS include its very fast incremental bounded model
checking algorithm and its flexible framework for experimenting with
novel analysis and abstraction ideas for invariant inference.
Drawbacks are its lack of support for certain program features
(e.g. multi-threading).
Table~\ref{tab:features} gives an overview of 2LS' features.
The remainder of the chapter is structured as follows:
\begin{itemize}
  \item Section~\ref{sec:approach} gives a high-level overview of the motivations and the theory behind 2LS' \emph{verification approach}.
  \item Section~\ref{sec:usecases} provides practical information on 2LS and
    \emph{how to} use it for selected kinds of verification tasks, such as
    proving functional properties and checking for memory safety, termination
    and overflows.
  \item Sections~\ref{sec:representation}--\ref{sec:features} describe the
    underlying theoretical concepts behind 2LS.
  \item Section~\ref{sec:architecture} provides details on the 2LS software
    project and its architecture.
  \item Section~\ref{sec:evaluation} reports data on experimental comparisons in the Software Verification Competition, in which 2LS has been participating since 2016.
\end{itemize}

\begin{table}[t]
  \caption{Features of 2LS.}
  \begin{tabular}{|p{0.25\textwidth}|p{0.7\textwidth}|}
    \hline
    Languages  & C, GOTO \\
    \hline
    Properties & \texttt{assert}, memory safety, overflow, division-by-zero,
                         termination\\
    \hline
    Environments  & Linux \\
    \hline
    Technologies used  & symbolic execution, bounded model checking \\
                       & k-induction \\
                       & abstract interpretation\\
                       & template-based predicate synthesis\\
                       & ranking functions, recurrent sets \\
                       & SAT and SMT solving \\
    \hline
    Other features & multiple input files, compilation and linking of entire projects into GOTO via \texttt{goto-cc}\\
    \hline
    Current strengths  & memory safety, floating point, loops \\
    \hline
    Current weaknesses & recursion, multi-threading \\
    \hline
  \end{tabular}
  \label{tab:features}
\end{table}


\section{Verification Approach} \label{sec:approach}



The software verification literature contains a wide range of
techniques which can be used to prove or disprove properties about
programs, each with their own strengths and weaknesses.  This presents
quite a challenge for non-expert users.
Often the choice of which tools to use and where to expend effort depends on
whether the properties to be verified are true or not---which is exactly what
they want to find out.
Hence, to build a robust and usable software verification system, it is
necessary to combine a variety of techniques.

One option would be to run a series of independent tools, in parallel
(as a portfolio, for example) or in some sequential order.  However,
this limits the information that can be exchanged between the
algorithms.

Another option would be to use monolithic algorithms such as
CEGAR~\cite{CGJ+00}, IMPACT~\cite{Mcm06}, or IC3/PDR~\cite{BM07,HB12}
which combine some of the \emph{ideas} of simpler systems. These are
difficult to implement well as their components interact in complex
and subtle ways.  Also, they require advanced solver features such as
interpolant generation that are not widely available for all theories
(bit-vectors, arrays, floating-point, etc.). 2LS takes a different approach to
the problem as explained below.

\subsection{$k$-induction and $k$-invariants (\kiki)}

2LS strives to build a verifier as a compound
with simple components and well-understood interaction.
These include:

\begin{description}
  \item[Bounded Model Checking]{Given sufficient time and resource, BMC
    will give counterexamples for all false safety properties,
    which are often of significant value for understanding the fault.  
    However, only a small proportion of true properties can
    be proven by BMC.}

  \item[$\mathbf{k}$-Induction]{Generalising Hoare logic's ideas of loop
   invariants, $k$-induction can prove true safety properties, and, in some
   cases, provide counterexamples to false ones.
   However, it requires inductive invariants, which can be expensive (in
   terms of user time, expertise, and maintenance).}

  \item[Abstract Interpretation]{The use of over-approximations makes
   it easy to compute invariants which allow many true
   propositions to be proven.  However false properties and
   true-but-not-provable properties may be indistinguishable.
   Tools may have limited support for a more complete analysis.}
\end{description}

2LS's \kiki algorithm~\cite{BJKS15} draws together a range of
well-known techniques and combines them in a novel way so that they
strengthen and reinforce each other.  The $k$-induction
technique~\cite{SSS00} uses syntactically restricted or simple
invariants (such as those generated by abstract interpretation) to
prove safety.  Bounded model checking~\cite{BCCZ99} allows us to test
$k$-induction failures to see if they are real counterexamples or, if
not, to build up a set of assumptions about system behaviour.
Template-based abstract interpretation is used for invariant
inference~\cite{SSM05,RSY04,GSV08} with unrolling producing
progressively stronger invariants, allowing the techniques to
strengthen each other.

\kiki unifies these techniques in a simple and elegant algorithm for integrated
invariant inference and counterexample generation.
The main loop of $\kiki$ is based
on incremental unwinding of the transition relation.
In the $k$-th iteration, the transition relation is unwound
  up to depth $k$ and a $k$-inductive invariant is inferred. The
invariant is then used to strengthen the program's safety property in
order to find a proof for the program's safety. In case that safety
cannot be proved, the algorithm checks whether the current unwinding is
sufficient to generate a counterexample. If this is not the case, the
unwinding~$k$ is incremented and another iteration starts.
The $\kiki$ algorithm is explained in more detail in Section
  \ref{sec:kiki}.
%

Internally, 2LS reduces program analysis problems expressed in second
order logic such as invariant or ranking function inference to
synthesis problems over templates. Hence, it reduces (an existential
fragment of) 2nd order Logic Solving to quantifier elimination in
first order logic. Loop unwinding as used by bounded model checking
and $k$-induction are considered as refinements of the original
program abstraction.
That way, 2LS provides an efficient implementation of \kiki based on
incremental loop unwinding, template-based predicate synthesis, and,
ultimately, incremental SAT solving.
These concepts, along with other important underlying concepts of 2LS
are explained in more detail throughout Sections~\ref{sec:representation},
\ref{sec:kiki}, and~\ref{sec:features}.




\section{Use Cases} \label{sec:usecases}

Before explaining the theoretical concepts behind 2LS, we give an overview of
2LS verification capabilities in the form of a tutorial.
2LS has a broad set of functionalities that allows the user
to solve a variety of verification problems.
We will use example programs, which can be found at \url{https://github.com/diffblue/2ls/tree/master/regression/book-examples}.

\subsection{Proving Functional Properties}

Functional properties can be specified in C programs with the help
of assertions, as provided by \texttt{assert.h}.
In order to state preconditions, the CPROVER framework provides
the \texttt{\_\_CPROVER\_assume(cond)} function.%
\footnote{In order to keep the code compilable, this function can be surrounded
by \texttt{\#ifdef \_CPROVER\_ ... \#endif}, which is defined when running
2LS.}

As a first example, let us consider the function in
Figure~\ref{fig:ex-numerical-data}, which extracts the authority part of
a URI given as a character string \texttt{uri} into the buffer
\texttt{authority}.
We would like to ensure that we do not copy beyond the length of the URI,
which is expressed by the assertion on line~9.

\begin{figure}[t]
\begin{lstlisting}
  void copy_authority(
    char *uri, int uri_length, int authority_start, char *authority) {
  __CPROVER_assume(0 < uri_length);
  __CPROVER_assume(0 < authority_start && authority_start < uri_length);
  int cp = authority_start;
  while (cp != uri_length - 1) {
    if (uri[cp] == '/')
      break;
    assert(cp < uri_length);
    authority[cp - authority_start] = uri[cp];
    ++cp;
  }
}
\end{lstlisting}
\caption{\texttt{uri.c}: numerical computations in a loop.}
\label{fig:ex-numerical-data}
\end{figure}

\begin{figure}[b]
\begin{lstlisting}
extern int __VERIFIER_nondet_int();

#include <stdlib.h>
#include <limits.h>

#define APPEND(l,i) {i->next=l; l=i;}

typedef struct node {
    struct node *next;
    int val;
} Node;

int main() {
    Node *l = NULL;
    int min = INT_MAX;
 
    while (__VERIFIER_nondet_int()) {
        Node *p = malloc(sizeof(*p));
        p->val = __VERIFIER_nondet_int();
        APPEND(l, p)

        if (min > p->val)
            min = p->val;
    }
    for (Node *i = l; i != NULL; i = i->next)
        assert(i->val >= min);
}
\end{lstlisting}
\caption{\texttt{min.c}: data stored in a dynamically allocated data structure.}
\label{fig:ex-heap-data}
\end{figure}

2LS provides various algorithms for verifying this property. For example:
\begin{itemize}
\item \texttt{2ls uri.c -{}-function copy\_authority -{}-havoc}\\
  2LS performs a 1-induction check using the invariant $\true$.
  This is not sufficient to prove the property.
  Hence, it returns \unknown.
\item \texttt{2ls uri.c -{}-function copy\_authority -{}-intervals}\\
  2LS performs a 1-induction check using the interval invariants.
  This is still not sufficient. We get the result \unknown.
\item \texttt{2ls uri.c -{}-function copy\_authority -{}-zones}\\
  Using the stronger zones abstract domain, we are able to prove the property.
  2LS outputs \ok.
\item \texttt{2ls uri.c -{}-function copy\_authority -{}-octagons}\\
  The even stronger octagons abstract domain allows us to prove the property,
  too. 2LS returns \ok.
\item \texttt{2ls uri.c -{}-function copy\_authority -{}-havoc -{}-k-induction}\\
  2LS performs $k$-induction. This is sufficient to prove the property, even
  though we do not infer additional invariants (the \texttt{havoc} option). 
  2LS reports \ok.
\end{itemize}

The example above dealt with numerical data. Let us now have a look at the
program in Figure~\ref{fig:ex-heap-data}. This program allocates a
singly-linked list of arbitrary length whose elements hold some integer value.
\footnote{The \texttt{\_\_VERIFIER\_nondet\_int} function is declared
  but does not have a definition. 2LS assumes that functions without
definition return a non-deterministic value.}
It then computes the \texttt{min}imum value held by the list elements
and checks whether the minimum has been computed correctly, i.e.,
it must be smaller than or equal to all values in the list.

2LS can prove this thanks to combined reasoning about the shape of
unbounded data structures and their contents at the same time.
Running
\begin{verbatim}
2ls min.c --heap --values-refine
\end{verbatim}
will return \ok{} for the property on line 26\footnote{The
\texttt{-{}-heap --values-refine} options employ a combination of shape and
numerical domains.}.
This means that the assertion is not violated by any list of any size.


\subsection{Finding Memory Safety Bugs}

Another class of properties that can be verified using 2LS is that of memory
safety properties. These include, e.g., array accesses within bounds, pointer
dereference safety, free safety, or absence of memory leaks. 2LS performs
verification of these properties by instrumenting the program with custom
assertions. The complete list of available instrumentations can be seen by
running 2LS with the \texttt{-{}-help} option.

We demonstrate verification of memory safety properties on two examples. First,
let us consider the program in Figure~\ref{fig:ex-array-bounds} which
transforms a two-dimensional array \texttt{matrix} into a single-dimensional
array \texttt{array}. We let the matrix dimensions to be random, but we add a
precondition on line 7 assuring that the number of matrix elements fits the
destination array.

\begin{figure}
\begin{lstlisting}
#define SIZE 1000

int main() 
    int **matrix;
    unsigned m = __VERIFIER_nondet_int();
    unsigned n = __VERIFIER_nondet_int();
    __CPROVER_assume(m <= SIZE && n <= SIZE && m * n <= SIZE);

    int array[SIZE];
    for (int row = 0; row < m; ++row) {
        for (int col = 0; col < n; ++col) {
            int index = row * m + col;
            array[index] = matrix[row][col];
        }
    }
    return 0;
}
\end{lstlisting}
\caption{\texttt{matrix\_to\_vector.c}: array bounds safety.}
\label{fig:ex-array-bounds}
\end{figure}

Running the verification with
\begin{verbatim}
2ls --bounds-check --k-induction matrix_to_vector.c
\end{verbatim}
will return \fail{} with an additional information about the
particular generated assertion that failed:
{
\begin{verbatim}
[main.2] array `array' upper bound in array[index]: FAILURE
\end{verbatim}}
\noindent This information says that \texttt{index} may be larger than the
array's upper bound (\texttt{SIZE}). Furthermore, 2LS is able to provide a
concrete counterexample in case it found a property violation. We may get this
by re-running 2LS with an additional \texttt{-{}-trace} option. An important
part of the obtained trace is shown in Figure~\ref{fig:ex-array-bounds-output}.
\begin{figure}
{
\begin{verbatim}
...

file main.c line 5 function main
  m=1000u

file main.c line 6 function main
  n=1u

...

file main.c line 10 function main
  row=1

file main.c line 11 function main
  col=0

file main.c line 12 function main
  index=1000
\end{verbatim}}
\caption{Counterexample for array bound safety violation for program in
Figure~\ref{fig:ex-array-bounds}.}
\label{fig:ex-array-bounds-output}
\end{figure}

Using the counterexample, we discover an error on line 12, which should
be:
\begin{lstlisting}[firstnumber=12, basicstyle=\ttfamily\footnotesize,
numberstyle=\scriptsize]
int index = row * n + col;
\end{lstlisting}

Running the verification again after correcting the error returns \ok,
which means that no more out-of-bounds array accesses are present in the
program.

For the second example of memory safety verification, we consider the program
containing a dynamic memory allocation shown in
Figure~\ref{fig:ex-free-safety}. The program allocates two memory objects
pointed by pointers \texttt{a} and \texttt{b} and linked together (the value of
the object pointed by \texttt{b} equals to \texttt{a} which points to a
random integer). Afterwards, both allocated objects are freed.  However, the
freeing order is incorrect since \texttt{b} is freed first and an attempt to
free the pointed object will likely fail due to an invalid dereference of a
dangling pointer. 2LS can be used to verify absence of pointer dereference and
of memory free errors by running
\begin{verbatim}
2ls --pointer-check --heap free_safety.c
\end{verbatim}
As expected, running this command on the shown program returns \fail{}
with an additional information about the error:
{
\begin{verbatim}
[main.11] dereference failure: deallocated dynamic object in *b: FAILURE
\end{verbatim}}

\begin{figure}
\begin{lstlisting}
int main() {
    int *a = malloc(sizeof(int));
    *a = __VERIFIER_nondet_int();

    int **b = malloc(sizeof(int *));
    *b = a;

    free(b);
    free(*b);
}
\end{lstlisting}
\caption{\texttt{free\_safety.c}: free safety verification.}
\label{fig:ex-free-safety}
\end{figure}

\subsection{Termination and Nontermination}

2LS can prove whether a program is terminating or
whether it may not terminate under certain input conditions.

\begin{figure}
\begin{lstlisting}
void bubble_sort(int *array, int size)
{
  __CPROVER_assume(size >= 0);
  int c, d, swap;

  for (c = 0; c < size - 1; c++)
  {
    for (d = 0; d < size - c - 1; d++)
    {
      if (array[d] > array[d + 1])
      {
        swap       = array[d];
        array[d]   = array[d + 1];
        array[d + 1] = swap;
      }
    }
  }
}
\end{lstlisting}
\caption{\texttt{bubble\_sort.c}: proving termination.}
\label{fig:ex-termination}
\end{figure}

For example, Figure~\ref{fig:ex-termination} shows
an implementation of the bubble sort algorithm.
We can check whether it terminates by running the following:
\begin{verbatim}
2ls bubble_sort.c --termination --function bubble_sort
\end{verbatim}
We get an output of the following sort:
{
\begin{verbatim}
...
Summary for function bubble_sort
...
termination argument:
($guard#25 && ... ==> (-1 * (d#phi25 - d#30) + 0 * ...) > 0) &&
($guard#23 && ... ==> (-1 * (c#phi23 - c#32) + 0 * ...) > 0)
...
** Termination: 
...
[bubble_sort]: yes
VERIFICATION SUCCESSFUL
\end{verbatim}}

2LS computes a termination argument in the form of a ranking function.
The termination argument in the output above is heavily simplified to
make it fit on this page. The actual termination argument output by 2LS
has more elaborate guard constraints and casts to enable correct bitvector
arithmetic without overflows.
An inspection of the termination argument says that \texttt{-c} is a
ranking function for the outer loop and \texttt{-d} is a ranking
function for the inner loop, which proves termination of the
\texttt{bubble\_sort} function.

\begin{figure}
\begin{lstlisting}
long sasum(long *sx, int n, int incx)
{
  int nincx = n * incx;
  long stemp = 0l;
  int i;
  for (i = 0; incx < 0 ? i >= nincx : i <= nincx; i += incx)
  {
    stemp += sx[i-1];
  }
  return stemp;
}
\end{lstlisting}
\caption{\texttt{sasum.c}
  (cf. \protect\url{http://www.netlib.org/clapack/cblas/sasum.c).}}
\label{fig:ex-non-termination}
\end{figure}

2LS can also check for non-termination. It will do this by finding
lasso-shaped executions. Figure~\ref{fig:ex-non-termination} shows
an array summation algorithm from a linear algebra package.
Running
\begin{verbatim}
2ls sasum.c --function sasum --nontermination --trace
\end{verbatim}
gives the output shown in Figure~\ref{fig:ex-non-termination-output}.
Inspecting the counterexample trace, we see that variable \texttt{i}
has the recurring value \texttt{0} on line 6, which implies
non-termination of the loop because the loop condition can never
become false. We can even see an example of input conditions
under which this happens: \texttt{n=2, incx=0}.

\begin{figure}
{
\begin{verbatim}
Unwinding (k=1)
...
Nonterminating program execution proved after 1 unwinding(s)
...

Counterexample:

file sasum.c line 1
  n=2
file sasum.c line 1
  incx=0
...
file sasum.c line 6 function sasum
  i=0
...
file sasum.c line 6 function sasum
  i=0

VERIFICATION FAILED
\end{verbatim}}
\caption{Output for the non-termination example.}
\label{fig:ex-non-termination-output}
\end{figure}


In practice, one can run \texttt{-{}-termination} and
\texttt{-{}-nontermination} in parallel and wait for the first to
report a conclusive result.

\subsection{Overflows}

Another property that can be verified using 2LS is safety from integer
overflows (and underflows). Similarly to memory safety verification, it is done
by instrumenting the analysed program with custom assertions.

As an illustration, we use the program in Figure~\ref{fig:ex-termination}. In
the program, the precondition \texttt{\_\_CPROVER\_assume(size >= 0)} on line 3
is necessary to avoid signed arithmetic underflows in computing \texttt{size -
1}, which would be an undefined behaviour in the C standards supported by 2LS.
We can verify this by running 2LS to check for overflows:

{
\begin{verbatim}
2ls  bubble_sort.c --signed-overflow-check --function bubble_sort
\end{verbatim}
}

\noindent Without the precondition, we would get:

{
\begin{verbatim}
...
[bubble_sort.1] arithmetic overflow on signed - in size - 1: FAILURE
...
VERIFICATION FAILED
\end{verbatim}}
\noindent which confirms the necessity of the precondition.

\section{Internal Program Representation} \label{sec:representation}

Being built upon the CPROVER
infrastructure~\cite{CKL04}, 2LS uses \emph{GOTO
  programs} as an intermediate representation.  In this language, any
non-linear control flow, such as if or switch-statements, loops, or
jumps, is translated to equivalent \emph{guarded goto} statements.
These statements are branch instructions that include
(optional) conditions.  CBMC generates one GOTO program per C function
found in the parse tree.  Furthermore, it adds a new main function
that first calls an initialisation function for global variables and
then calls the original program entry function.

Similarly to CBMC, 2LS performs a light-weight static analysis to resolve
function pointers to a case split over all candidate functions, resulting in
a static call graph.  Furthermore, assertions guarding against invalid
pointer operations or memory leaks are inserted. In addition, 2LS uses local
constant propagation and expression simplification to increase efficiency.

After running the mentioned transformations, 2LS performs a static analysis to
derive data flow equations for each function of the GOTO program.  The result
is a \emph{static single assignment (SSA) form} in which loops have been cut
at the back edges to the loop head.  The effect of these cuts is havocking of
the variables modified in the loop at the loop head.  This SSA is hence an
over-approximation of the GOTO program. Subsequently, 2LS refines this
over-approximation by computing invariants.

Since this SSA representation is crucial for the verification approach
of 2LS, we give its detailed description in
Section~\ref{sec:ssa}. Prior to that, we introduce our memory model
that is used to represent dynamically allocated objects in our SSA
form (Section~\ref{sec:memmodel}). At last, we give
an insight into our loop unwinding approach, which is an imporant step
of the $\kiki$ algorithm, in Section~\ref{sec:unwinding}.

\subsection{Memory Model} \label{sec:memmodel}

We now describe a memory model that we use to represent all program
memory~\cite{MHSV18}. Our model is object-based, and we distinguish objects
allocated \emph{statically} (i.e., variables on the stack and global variables)
and \emph{dynamically} (i.e., on the heap). 

\subsubsection{Static memory objects}

In our approach, we work with non-recursive programs with all functions
inlined. Therefore, we do not need to consider the stack and the set $\var$ of
static memory objects corresponds to the set of all program variables. Each
variable is uniquely identified by its name.

For convenience, we define subsets of $\var$ that correspond to sets of
variables of a chosen type:
\begin{itemize}
  \item $\nvar$ is the set of all variables of a numeric type (integer or
    floating point).
  \item $\pvar$ is the set of all variables of a pointer type.
  \item $\svar$ is the set of all variables of a structure type. Each structure
    type defines a set of named \emph{fields}, each of them having its own
    type. We use $\fld$ to denote the set of all fields used in the analysed
    program.  Similarly to the variables, we let $\nfld, \pfld \subseteq \fld$
    be the sets of all fields of numerical and pointer types, respectively. In
    order to express access to individual fields of a structure-typed variable,
    we use the ``dot'' notation as is common in C.
\end{itemize}
We assume $\nvar$, $\pvar$, and $\svar$ are pairwise disjoint.

\subsubsection{Dynamic memory objects}

To represent dynamic memory objects (i.e., those allocated using
\texttt{malloc} or some of its variants), we use \emph{abstract dynamic
objects}. An abstract dynamic object represents a set of concrete dynamic
objects allocated by the same \texttt{malloc} call. We refer to a
\texttt{malloc} call at a program location $i$ as to an \emph{allocation site}
$i$. 

Generally, a single abstract dynamic object is not sufficient to represent all
concrete objects allocated by a single \texttt{malloc} call. This is due to the
fact that the analysed program may use several concrete objects allocated at
the same allocation site at the same time. If such objects are, e.g., compared,
our memory model must allow us to distinguish them. This can be done either by
concretisation on demand (as is common in many memory models) or by
pre-materialisation of a sufficient number of objects at the beginning of the
analysis. Since our approach uses a single formula to represent the analysed
program and leverages on small incremental changes of the formula during the
analysis, we use the latter approach. 

Therefore, we use the set $\dynobjset{i} = \{\dynobj{i}{k} \mid 1 \leq k \leq
n_i\}$ of abstract objects to represent all concrete objects allocated at the
allocation site $i$. The number $n_i$ of necessary objects is determined for
each allocation site using an approach described later in this section. The set
of all dynamic objects of the analysed program is then defined as $\dobj =
\cup_i \dynobjset{i}$. Together with the set of all static objects, we define
$\obj = \var \cup \dobj$ to be the set of all memory objects of our program
abstraction. We require $\var \cap \dobj = \emptyset$ and $\dynobjset{i} \cap
\dynobjset{j} = \emptyset$ for $i \neq j$.

Similarly to static objects, we denote $\ndobj$, $\pdobj$, and $\sdobj$ the
sets of dynamic objects of a numerical, pointer, and structure type,
respectively. Fields of structure-typed dynamic objects, i.e., elements of the
set $\sdobj \times \fld$, represent abstractions of the appropriate field of
all represented concrete objects. Using the above sets, we may define the set
$\num$ of all \emph{numerical objects} in the program as:
\begin{equation}
  \num = \nvar \cup \ndobj \cup ((\svar \cup \sdobj) \times \nfld).
\end{equation}
Analogically, we define the set $\ptr$ of all \emph{pointers} as:
\begin{equation}\label{eq:pointers}
  \ptr = \pvar \cup \pdobj \cup ((\svar \cup \sdobj) \times \pfld).
\end{equation}
Pointers can be assigned addresses of objects. Since we do not support pointer
arithmetic, only \emph{symbolic addresses} and a special address $\nullptr$ are
considered. We use the operator \& to get the address of both static and
dynamic objects. For abstract dynamic objects, the symbolic address is an
abstraction of symbolic addresses of all represented concrete objects. We
define the set $\addr$ of all addresses in the program as:
\begin{equation}
  \addr = \{\&o \mid o \in \obj\} \cup \{\nullptr\}.
\end{equation}

\subsubsection*{Dynamic Object Pre-Materialisation}

Above, we mentioned that, for each allocation site $i$, we represent (a
potentially infinite number of) all objects allocated at $i$ by a finite number
$n_i$ of abstract dynamic objects. In order for this abstraction to be sound,
it is sufficient that this number is equal to the number of distinct concrete
objects, allocated at $i$, that may be simultaneously pointed to at any
location of the analysed program.

In order to compute this number, we first perform a standard static
\emph{may-alias analysis}. This analysis determines, for each program location
$j$, the set $P_i^j$ of all \emph{pointer expressions} used in the program,
that may point to some object allocated at $i$. Here, pointer expressions may
be one of the following:
\begin{itemize}
  \item Pointer variables.
  \item Dereferences of pointers to pointers. These correspond to pointer-typed
    dynamic objects.
  \item Pointer fields of structure-typed variables.
  \item Dereferences of pointers to structures followed by an access to a
    pointer-typed field. These correspond to pointer-typed fields of dynamic
    objects.  Here, we use the C notation based on an arrow (e.g.,
      $p\!\!\rightarrow\!\!n$ to express a dereference of $p$ followed by an
      access to the
    field $n$ of the pointed object).
\end{itemize}
For simplicity, we assume that all chained dereferences of forms **$p$ or
$p\!\!\rightarrow\!\!f_1\!\!\rightarrow\!\!f_2$ are split into multiple
dereferences using an
intermediate pointer variable which is added to $\pvar$.

Next, we compute the must alias relation $\sim_j$ over the set of all pointer
expressions. For each pair of pointer expressions $p$ and $q$ and for each
program location $j$, $p \sim_j q$ iff $p$ and $q$ must point to the same
concrete object (i.e., they must alias) at $j$.

Finally, we partition each computed $P^j_i$ into equivalence classes by
$\sim_j$, and the number $n_i$ is given by the maximal number of such classes
for any $j$.


\subsection{The Static Single Assignment Form} \label{sec:ssa}

Program verification in 2LS is based on generating program abstractions using a
solver. In order to simplify generation of a formula representing the program
semantics, 2LS uses the \emph{static single assignment form} (SSA) to represent
programs. SSA is a standard program representation used in bounded model
checking or symbolic execution tools. We use common concepts of
SSA---introducing a fresh copy (version) $x_i$ of each variable $x$ at program
location $i$ in case $x$ is assigned to at $i$, using the last version of $x$
whenever $x$ is read, and introducing a \emph{phi} variable $\ssaxphi{i}$ at a
program join point $i$ in case different versions of $x$ come from the joined
program branches. For an acyclic program, SSA is a formula that represents
exactly the post condition of running the code.

We extend the traditional SSA by three new concepts: (1) over-approximation of
loops in order to make the SSA acyclic~\cite{BJKS15}, (2) a special encoding of
the control-flow~\cite{BJKS15}, and (3) a special representation of
memory-manipulating operations that comply with our memory model introduced
in Section~\ref{sec:memmodel}~\cite{MHSV18}. In the rest of this section, we
describe these extensions.

\subsubsection{Over-Approximation of Loops}

In order to be able to use a solver for reasoning about program abstractions,
we extend the SSA by \emph{over-approximating} the effect of loops. As was said
above, the value of a variable $x$ is represented at the loop head by a phi
variable $\ssaxphi{i}$ joining the value of $x$ from before the loop and from
the end of the loop body (here, we assume that all paths in the loop join
before its end, and the same holds for the paths before the loop). However,
instead of using the version of $x$ from the loop end, it is replaced by a
\emph{free loop-back variable} $\ssaxlb{}$. This way, the SSA remains acyclic,
and, since the value of $\ssaxlb{}$ is initially unconstrained, the effect of
the loop is over-approximated. To improve precision, the value of $\ssaxlb{}$
can be later constrained using a \emph{loop invariant} that will be inferred
during the analysis. A loop invariant is a property that holds at the end of
the loop body, after any iteration and can be therefore assumed to hold on the
loop-back variable.

\begin{figure}[h!]
\begin{center}
\begin{subfigure}[b]{0.2\textwidth}
\begin{lstlisting}[xleftmargin=1.5em]
unsigned x = 0;

while (x < 10)
{
    ++x;
}
\end{lstlisting}
\caption{A loop in C.}
\end{subfigure}
\hspace{2em}
\begin{subfigure}[b]{0.4\textwidth}
\begin{tikzpicture}
    \node [] (i) at (0,0) {before the loop};
    \node [] (i2) at (0,-0.5) {$\ssax{0} = 0$};
    \node [] (1) at (-1.9, 0.01) {\scriptsize 1:};
    \node [] (h) at (0,-2) {\bf loop head};
    \node [] (3) at (-1.9, -1.98) {\bf\scriptsize 3:};
    \node [] (h2) at (0,-2.5) {$\ssaxphi{3} = \ssagls{6} \;?\; \ssaxlb{6} \;:\; \ssax{0}$};
    \node [] (b) at (0,-3.5) {loop body};
    \node [] (4) at (-1.9, -3.49) {\scriptsize 4:};
    \node [] (b2) at (0,-4) {$\ssax{5} = \ssaxphi{3} + 1$};
    \node [] (5) at (-1.9, -3.99) {\scriptsize 5:};
    \node [] (e) at (0,-5) {end of the loop body};
    \node [] (6) at (-1.9, -4.99) {\scriptsize 6:};
    \node [] (a) at (0,-6) {after the loop};

    \path [->] (i2) edge (h);
    \path [->] (h2) edge (b);
    \path [->] (b2) edge (e);
    \path [->] (h2.south west) edge[bend right=60, looseness=2] (a.west);
    \path [->] (e.east) edge[bend right=80, looseness=2] node[right]{$\ssaxlb{6}$} (h.east);
\end{tikzpicture}
\caption{Encoding of the loop in the SSA form.}
\end{subfigure}
\caption{Conversion of loops in the SSA form used in 2LS.}
\label{fig:ssa_loops}
\end{center}
\end{figure}
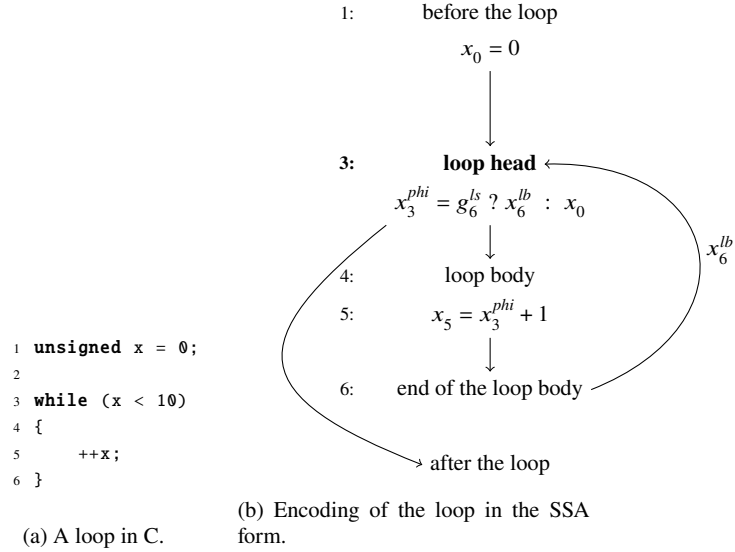

For a better illustration, we give an example of this SSA extension.
Figure~\ref{fig:ssa_loops} shows a simple loop in C and its corresponding SSA.
Instead of using $\ssax{5}$ in the phi variable, a fresh variable $\ssaxlb{6}$
is introduced. Moreover, the join in the phi node is driven by a free Boolean
variable $\ssagls{6}$ (a so-called \emph{loop-select guard}) modelling a
non-deterministic choice between $\ssaxlb{6}$ and $\ssax{0}$.

\subsubsection{Encoding the Control-Flow}

In 2LS, the program is represented by a single monolithic formula. It is
therefore needed that the formula encodes control-flow information. This is
achieved using so-called \emph{guard} variables that track the reachability
information for each program location. In particular, for each program location
$i$, we introduce a Boolean variable $\ssag{i}$ whose value encodes whether
$i$ is reachable. For example, in Figure~\ref{fig:ssa_loops}, a guard
$\ssag{5}$ encoding reachability of the loop body would have the value:
\begin{equation}
  \ssag{5} = \ssaxphi{3} < 10.
\end{equation}

\subsubsection{Representation of Memory-Manipulating
Operations}\label{sec:memops}

\subsubsection*{Dynamic Memory Allocation}

As said in Section~\ref{sec:memmodel}, all concrete objects allocated by a
single \texttt{malloc} call at a program location $i$ are abstracted by a set
of abstract dynamic objects $\dynobjset{i}$. In the SSA form, we represent such
a call by a non-deterministic choice among objects from $\dynobjset{i}$. A
program assignment \verb#p = malloc(...)# is therefore transformed into the
formula

\begin{equation}
\ssa{p}{i} = \ssagosij{i}{1} \,?\, \&\dynobj{i}{1} :
(\ssagosij{i}{2} \,?\, \&\dynobj{i}{2} : (\ldots (\ssagosij{i}{n_i-1}
\,?\, \&\dynobj{i}{n_i-1} : \&\dynobj{i}{n_i})))
\end{equation}
where $\ssagosij{i}{j}$, $1 \leq j < n_i$, are free Boolean variables, so-called
\emph{object-select guards}.

\subsubsection*{Reading through Dereferenced Pointers}

We now describe encoding of a pointer dereference appearing on the right-hand
side of an assignment or in a condition (i.e., in an R-expression).  Prior to
generating the SSA, we perform a static \emph{may-points-to analysis} which
over-approximates---for each program location $i$ and for each pointer $p \in
\ptr$---the set of all objects from $\obj$ that $p$ may point to at $i$. A
dereference of $p$ at $i$ is then represented by a choice among the pointed
objects.

Moreover, to simplify the representation and to improve precision (for reasons
explained below), we also introduce so-called \emph{dereference variables}.
Let $*p$ be an R-expression that appears at a program location $i$ and let us
assume that $p$ may point to a set of objects $O \subseteq \obj$. We replace
$*p$ by a fresh variable $\drfvar{p}{i}$, and we define its value as follows:
\begin{equation}\label{eq:mem_read}
\begin{split}
&\bigwedge_{o \in O} \ssa{p}{j} = \&o \Longrightarrow \drfvar{p}{i} = \ssa{o}{k}
\;\wedge\\
&\left(\bigwedge_{o \in O} \ssa{p}{j} \neq \&o\right) \Longrightarrow 
  \drfvar{p}{i} = o_\perp
\end{split}
\end{equation}
where $\ssa{p}{j}$ and $\ssa{o}{k}$ are the relevant versions of the variables $p$
and $o$, respectively, at the program location $i$, and $o_\perp$ is a special
``unknown object'' representing the result of a dereference of an unknown or
invalid ($\nullptr$) pointer.

Informally, \eqref{eq:mem_read} expresses the fact that $\drfvar{p}{i}$
equals the value of $o$ at $i$ in case that $p$ points to $o$ at $i$, and
it equals the value of the unknown object otherwise.

\subsubsection*{Writing through Dereferenced Pointers}

Similarly to the operation of reading, we introduce an SSA encoding for the
operation of writing into memory using a pointer dereference. Again, we leverage
on a may-points-to analysis described above, and we introduce special
\emph{dereference variables}.  Let us have an assignment $*p = v$ at program
location $i$ and let us assume that $p$ may point to a set of objects $O
\subseteq \obj$ at the entry to $i$.  This assignment is replaced by an
equality:
\begin{equation}
  \drfvar{p}{i} = \ssa{v}{l}.
\end{equation}
where $\ssa{v}{l}$ is the valid version of $v$ at $i$. The dereference variable
$\drfvar{p}{i}$ is then used to update the value of the referenced object. This
is done using the formula:
\begin{equation}
\bigwedge_{o \in O} \ssa{o}{i} =
(\ssa{p}{j} = \&o) \,?\, \drfvar{p}{i} : \ssa{o}{k}
\end{equation}
where $\ssa{p}{j}$, $\ssa{o}{k}$ are relevant versions of the variables $p$ and
$o$, respectively, at program location $i$.

In other words, this formula expresses the fact that an object $o$ is assigned
the value of $v$ in the case when $p$ points to $o$, and it keeps its original
value otherwise. As mentioned above, usage of dereference variables may improve
precision of the representation. This happens especially when we write into an
abstract object through some pointer and afterwards read through the same
pointer without changing its value nor the value of the pointed object in
between. In such cases, we may reuse the same dereference variable which ensures
that we get the same value that was written, which may not happen otherwise
since we read from an abstract object representing a number of concrete objects. 






\subsection{Structural Transformations} \label{sec:unwinding}


The \kiki algorithm uses loop unwinding to refine the
control flow structure of the program.
There are several possible strategies of how to
perform this unwinding.
We have experimented in \cbmc~\cite{SKB+17}
with an unwinding strategy that follows the control flow
and unwinds loops one by one and performs property checking
after each unwinding.
With such a strategy, each loop is incrementally unwound until all
paths exit the loop or until a maximum depth $k$ is reached. We~can
detect that a loop is fully unwound at unwinding $j<k$ if every state
reached at unwinding $j$ does not satisfy the loop condition.  After
a loop has been unwound, \cbmc continues to the next loop. This
procedure is repeated until all loops have been unwound or a bug has
been found.  Recursive function calls are treated similarly.

\begin{figure}
\centering
   \begin{subfigure}{0.13\textwidth} 
    \subcaptionbox{An example of a program with three loops.}[0.8\textwidth]{
\small
~\\[9.3ex]
\begin{tikzpicture}[>=stealth,x=0.75cm,y=0.75cm]
\node[draw,circle](n0) at (0,9.5) {0};
\node[draw,circle](n1) at (0,8.5) {1};
\coordinate(p1l) at (-1,8.5);
\coordinate(p1r) at (1,8.5);
\node[draw,circle](n2) at (0,7.5) {2};
\coordinate(p2l) at (-0.5,7.5);
\coordinate(p2r) at (0.5,7.5);
\node[draw,circle](n3) at (0,6.5) {3};
\node[draw,circle](n4) at (0,5.5) {4};
\coordinate(p4r) at (0.5,5.5);
\coordinate(p5l) at (-0.5,4.75);
\coordinate(p5) at (0,4.75);
\coordinate(p5r) at (1,4);
\node[draw,circle](n5) at (0,4) {5};
\coordinate(p6l) at (-1,3.25);
\coordinate(p6) at (0,3.25);
\node[draw,circle](n6) at (0,2.5) {6};
\coordinate(p6r) at (1,2.5);
\node[draw,circle](n7) at (0,1.5) {7};
\coordinate(p7r) at (1,1.5);
\coordinate(p7l) at (-1,2.5);
\coordinate(p8l) at (-1,0.75);
\coordinate(p8) at (0,0.75);
\node[draw,circle](n8) at (0,0) {8};
\draw[->] (n0) -- (n1);
\draw[->] (n1) -- (n2);
\draw[->] (n2) -- (n3);
\draw[->] (n3) -- (n4);
\draw[->] (n6) -- (n7);
\draw[->] (n1) -- (p1l) -- (p6l) --(p6) --(n6);
\draw[->] (n4) -- (p4r) -- (p2r) -- (n2);
\draw[->] (n2) -- (p2l) -- (p5l) -- (p5) -- (n5);
\draw[->] (n5) -- (p5r) -- (p1r) -- (n1);
\draw[->] (n6) -- (p7l) -- (p8l) -- (p8) -- (n8);
\draw[->] (n7) -- (p7r) -- (p6r) -- (n6);
\end{tikzpicture}
     }
   \end{subfigure}
   \begin{subfigure}{0.47\textwidth} 
    \subcaptionbox{Incremental unwinding strategy of
      \textsc{Cbmc}.}[0.8\textwidth]{
\small
\vspace*{24ex}
\begin{tikzpicture}[>=stealth,x=0.75cm,y=0.75cm]
\node(b0) at (0,10.5) {$\overbrace{\hspace*{2em}}^{\Phi(1)}$};
\node(b1) at (1.5,10.5) {$\overbrace{\hspace*{4em}}^{\Phi(2)}$};
\node(b2) at (4.25,10.5) {$\overbrace{\hspace*{8em}}^{\Phi(m)}$};
\node(b3) at (6.5,10.5) {\large\ldots};
\node(n00) at (6.5,9.5) {\large\ldots};
\node[draw,circle](n0) at (0,9.5) {0};
\node[draw,circle](n1) at (0,8.5) {1};
\node[draw,circle](n2) at (0,7.5) {2};
\node[draw,circle](n3) at (0,6.5) {3};
\node[draw,circle](n4) at (0,5.5) {4};
\draw[->] (n0) -- (n1);
\draw[->] (n1) -- (n2);
\draw[->] (n2) -- (n3);
\draw[->] (n3) -- (n4);
\node[draw,circle](n0) at (1,9.5) {0};
\node[draw,circle](n1) at (1,8.5) {1};
\node[draw,circle](n2) at (1,7.5) {2};
\node[draw,circle](n3) at (1,6.5) {3};
\node[draw,circle](n4) at (1,5.5) {4};
\coordinate(p4r) at (1.5,5.5);
\coordinate(p22l) at (1.5,8);
\coordinate(p22) at (2,8);
\node[draw,circle](n22) at (2,7.5) {2};
\node[draw,circle](n33) at (2,6.5) {3};
\node[draw,circle](n44) at (2,5.5) {4};
\draw[->] (n0) -- (n1);
\draw[->] (n1) -- (n2);
\draw[->] (n2) -- (n3);
\draw[->] (n3) -- (n4);
\draw[->] (n4) -- (p4r) -- (p22l) -- (p22) -- (n22);
\draw[->] (n22) -- (n33);
\draw[->] (n33) -- (n44);
\node[draw,circle](n0) at (3,9.5) {0};
\node[draw,circle](n1) at (3,8.5) {1};
\node[draw,circle](n2) at (3,7.5) {2};
\coordinate(p2l) at (2.6,7.5);
\node[draw,circle](n3) at (3,6.5) {3};
\node[draw,circle](n4) at (3,5.5) {4};
\coordinate(p4r) at (3.4,5.5);
\coordinate(p5l) at (2.6,4.75);
\coordinate(p22r) at (3.4,8);
\coordinate(p22) at (4,8);
\coordinate(p22l) at (3.6,7.5);
\node[draw,circle](n22) at (4,7.5) {2};
\node[draw,circle](n33) at (4,6.5) {3};
\node[draw,circle](n44) at (4,5.5) {4};
\coordinate(p44r) at (4.4,5.5);
\coordinate(p55l) at (3.6,4.75);
\coordinate(p222r) at (4.4,8);
\coordinate(p222) at (5.5,8);
\coordinate(p222l) at (5.1,7.5);
\node[draw,circle](n222) at (5.5,7.5) {2};
\coordinate(p555l) at (5.1,4.75);
\coordinate(p555r) at (4.4,4.75);
\coordinate(p555) at (5.5,4.75);
\node[draw,circle](n555) at (5.5,4) {5};
\draw[->] (n0) -- (n1);
\draw[->] (n1) -- (n2);
\draw[->] (n2) -- (n3);
\draw[->] (n3) -- (n4);
\draw[->] (n4) -- (p4r) -- (p22r) -- (p22) -- (n22);
\draw[->] (n22) -- (n33);
\draw[->] (n33) -- (n44);
\draw (n44) -- (p44r) -- (p222r);
\draw[dashed] (p222r) -- (p222);
\draw[->] (p222) -- (n222);
\draw[->] (n222) -- (p222l) -- (p555l) -- (p555) -- (n555);
\draw (n22) -- (p22l) -- (p55l) -- (p555r);
\draw[dashed] (p555r) -- (p555l);
\draw (n2) -- (p2l) -- (p5l) -- (p55l);
\end{tikzpicture}
     }
   \end{subfigure}
   \begin{subfigure}{0.38\textwidth} 
    \subcaptionbox{Incremental unwinding strategy of 2LS.}[0.8\textwidth]{
\small
\begin{tikzpicture}[>=stealth,x=0.75cm,y=0.75cm]
\node(b0) at (0,10.5) {$\overbrace{\hspace*{2em}}^{\Phi(1)}$};
\node(b1) at (2.75,10.5) {$\overbrace{\hspace*{11em}}^{\Phi(2)}$};
\node(b2) at (5.5,10.5) {\large\ldots};
\node(n00) at (5.5,9.5) {\large\ldots};
\node[draw,circle](n0) at (0,9.5) {0};
\node[draw,circle](n1) at (0,8.5) {1};
\coordinate(p1l) at (-0.75,8.5);
\coordinate(p1r) at (1,8.5);
\node[draw,circle](n2) at (0,7.5) {2};
\coordinate(p2l) at (-0.5,7.5);
\coordinate(p2r) at (0.5,7.5);
\node[draw,circle](n3) at (0,6.5) {3};
\node[draw,circle](n4) at (0,5.5) {4};
\coordinate(p4r) at (0.5,5.5);
\coordinate(p5l) at (-0.5,4.75);
\coordinate(p5) at (0,4.75);
\coordinate(p5r) at (1,4);
\node[draw,circle](n5) at (0,4) {5};
\coordinate(p6l) at (-0.75,3.25);
\coordinate(p6) at (0,3.25);
\node[draw,circle](n6) at (0,2.5) {6};
\coordinate(p6r) at (1,2.5);
\node[draw,circle](n7) at (0,1.5) {7};
\coordinate(p7r) at (1,1.5);
\coordinate(p7l) at (-0.75,2.5);
\coordinate(p8l) at (-0.75,0.75);
\coordinate(p8) at (0,0.75);
\node[draw,circle](n8) at (0,0) {8};
\draw[->] (n0) -- (n1);
\draw[->] (n1) -- (n2);
\draw[->] (n2) -- (n3);
\draw[->] (n3) -- (n4);
\draw[->] (n6) -- (n7);
\draw[->] (n1) -- (p1l) -- (p6l) --(p6) --(n6);
\draw[->] (n2) -- (p2l) -- (p5l) -- (p5) -- (n5);
\draw[->] (n6) -- (p7l) -- (p8l) -- (p8) -- (n8);
\node[draw,circle](n0) at (1.5,9.5) {0};
\node[draw,circle](n1) at (1.5,8.5) {1};
\coordinate(p1l) at (0.75,8.5);
\coordinate(p1r) at (2.5,8.5);
\node[draw,circle](n2) at (1.5,7.5) {2};
\coordinate(p2l) at (1,7.5);
\coordinate(p2r) at (2,7.5);
\node[draw,circle](n3) at (1.5,6.5) {3};
\node[draw,circle](n4) at (1.5,5.5) {4};
\node[draw,circle](n22) at (2.5,7.5) {2};
\coordinate(p4r) at (1.9,5.5);
\coordinate(p22r) at (1.9,8);
\coordinate(p22) at (2.5,8);
\coordinate(p22l) at (2.1,7.5);
\node[draw,circle](n33) at (2.5,6.5) {3};
\node[draw,circle](n44) at (2.5,5.5) {4};
\coordinate(p5l) at (1,4.75);
\coordinate(p5) at (1.5,4.75);
\coordinate(p5r) at (2.1,4.75);
\node[draw,circle](n5) at (1.5,4) {5};
\coordinate(p6l) at (0.75,3.25);
\coordinate(p6) at (1.5,3.25);
\node[draw,circle](n6) at (1.5,2.5) {6};
\node[draw,circle](n7) at (1.5,1.5) {7};
\node[draw,circle](n66) at (2.5,2.5) {6};
\coordinate(p66r) at (1.9,3);
\coordinate(p66) at (2.5,3);
\coordinate(p66l) at (2.1,2.5);
\node[draw,circle](n77) at (2.5,1.5) {7};
\coordinate(p8r) at (2.1,0.75);
\coordinate(p7r) at (1.9,1.5);
\coordinate(p7l) at (0.75,2.5);
\coordinate(p8l) at (0.75,0.75);
\coordinate(p8) at (1.5,0.75);
\node[draw,circle](n8) at (1.5,0) {8};
\draw[->] (n0) -- (n1);
\draw[->] (n1) -- (n2);
\draw[->] (n2) -- (n3);
\draw[->] (n3) -- (n4);
\draw[->] (n22) -- (n33);
\draw[->] (n33) -- (n44);
\draw[->] (n4) -- (p4r) -- (p22r) -- (p22) -- (n22);
\draw (n22) -- (p22l) -- (p5r) -- (p5);
\draw[->] (n6) -- (n7);
\draw[->] (n66) -- (n77);
\draw[->] (n1) -- (p1l) -- (p6l) --(p6) --(n6);
\draw[->] (n2) -- (p2l) -- (p5l) -- (p5) -- (n5);
\draw[->] (n6) -- (p7l) -- (p8l) -- (p8) -- (n8);
\draw (n66) -- (p66l) -- (p8r) -- (p8);
\draw[->] (n7) -- (p7r) -- (p66r) -- (p66) -- (n66);
\node[draw,circle](n11) at (3.5,8.5) {1};
\coordinate(p11l) at (2.9,9);
\coordinate(p11) at (3.5,9);
\coordinate(p11r) at (3,8.5);
\coordinate(p6r) at (3,3.25);
\node[draw,circle](n222) at (3.5,7.5) {2};
\coordinate(p222l) at (3.1,7.5);
\coordinate(p222r) at (4,7.5);
\coordinate(p55r) at (2.9,4);
\node[draw,circle](n333) at (3.5,6.5) {3};
\node[draw,circle](n444) at (3.5,5.5) {4};
\node[draw,circle](n2222) at (4.5,7.5) {2};
\coordinate(p444r) at (3.9,5.5);
\coordinate(p2222r) at (3.9,8);
\coordinate(p2222) at (4.5,8);
\coordinate(p2222l) at (4.1,7.5);
\node[draw,circle](n3333) at (4.5,6.5) {3};
\node[draw,circle](n4444) at (4.5,5.5) {4};
\coordinate(p555l) at (3.1,4.75);
\coordinate(p555) at (3.5,4.75);
\coordinate(p555r) at (4.1,4.75);
\node[draw,circle](n555) at (3.5,4) {5};
\draw[->] (n11) -- (n222);
\draw[->] (n222) -- (n333);
\draw[->] (n333) -- (n444);
\draw[->] (n2222) -- (n3333);
\draw[->] (n3333) -- (n4444);
\draw[->] (n444) -- (p444r) -- (p2222r) -- (p2222) -- (n2222);
\draw (n2222) -- (p2222l) -- (p555r) -- (p555);
\draw[->] (n222) -- (p222l) -- (p555l) -- (p555) -- (n555);
\draw[->] (n5) -- (p55r) -- (p11l) -- (p11) -- (n11);
\draw (n11) -- (p11r) -- (p6r) -- (p6);
\end{tikzpicture}
     }
   \end{subfigure}
\caption{Incremental unwinding strategies.} \label{fig:loopunwinding}
\end{figure}
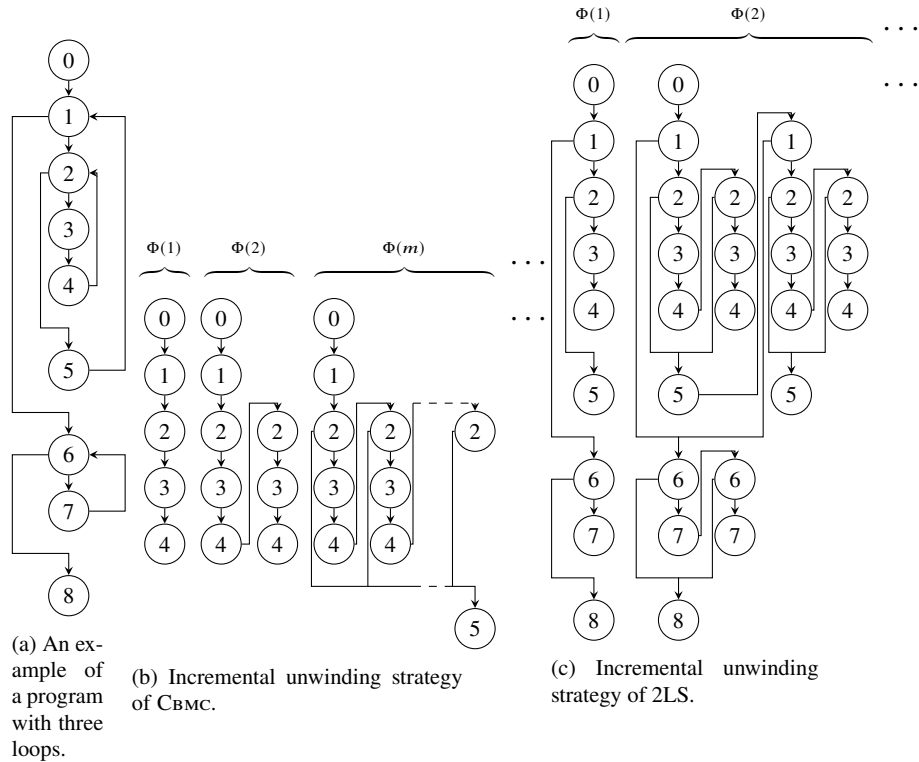

Consider the control flow graph (CFG) in Fig.~\ref{fig:loopunwinding}(a).  The
unwinding strategy is illustrated for this CFG in
Fig.~\ref{fig:loopunwinding}(b).  The program has three loops with loop
heads~1, 2, and 6 (2 is nested inside 1).  The symbolic execution that
generates the incremental BMC formula $\Phi(k)$ (see
Section~\ref{sec:incrbmc}) traverses the CFG and stops each time when it
encounters an edge in the CFG that returns to a loop head (a so-called
\emph{back-edge}).  Fig.~\ref{fig:loopunwinding}(b) shows three snapshots of
the partially unwound CFG that correspond to the parts of the program
considered by instances of the incremental BMC formula $\Phi(k)$ for
$k=1,2,m$.  We write $\Phi(1)$ for the formula up to the first back-edge
encountered that returns to the loop head of the inner loop (2).  Formula
$\Phi(2)$ extends $\Phi(1)$ by one further unwinding of the inner loop. 
Assume that $m$ is the maximum number of unwindings of the inner loop, then
$\Phi(m)$ shows the extension of the formula to the case where the inner
loop has been unwound up to this maximum number within the first iteration
of the outer loop (with loop head 1).  Formula $\Phi(m+1)$ will then extend
$\Phi(m)$ by a first unwinding of the inner loop (up to program location~4)
for the second iteration of the outer loop.  This process continues until a
failed assertion or the end of the program (8) is reached.

We realised that this approach is not very efficient. Hence,
we implement a different approach in 2LS~\cite{SK16,BJKS15}.
2LS unwinds \emph{all} loops $k$ times and incrementally adds the
$(k+1)\text{th}$ for \emph{all} loops instead of unwinding only the
first loop encountered until it has been fully unwound.

We~illustrate this unwinding strategy in Fig.~\ref{fig:loopunwinding}(c), which
shows the first two partial unwindings of the CFG in
Fig.~\ref{fig:loopunwinding}(a) that correspond to $\Phi(1)$ and $\Phi(2)$,
respectively.  Formula $\Phi(1)$ consists of one unwinding (up to, but not
including the back-edge) for the loops 1, 2, and~6.  Formula $\Phi(2)$ then
adds another unwinding to each loop.  Note that we have two times two
unwindings of the inner loop (with loop head 2) now, two for each unwinding
of the outer loop (loop head 1).

Structurally, this unwinding strategy is the same as the one that we
use in non-incremental \cbmc when calling it with fixed values for $k$.

2LS performs loop unwinding on the SSA form augmented with some meta-information
on the loop structure and hierarchy.
%
%
%
%
The unwinding that we perform is incremental, in the sense that we incrementally
extend the SSA when increasing $k$ to $k+1$.
The problematic part is the merging of loop exits (nodes 5, 6, and 8 in
Fig.~\ref{fig:loopunwinding}(c)).
We have to account for the case of ``value is merged from an unwinding that has
not been added yet'', which makes the construction of the formula non-monotonic.
Indeed, one has to effectively \emph{remove} the disjunction describing each
such merge, which represents one conjunct in the SSA form, and replace it by a
larger disjunction reflecting the addition of a new incoming branch of the
concerned merge obtained by an additional unwinding.
Removing parts of the formula, however, is not supported by SAT solvers.
Fortunately, this problem can be solved with the help of \emph{solving under
assumptions} as we will explain in Section~\ref{sec:sat}.
This is an essential ingredient for performance as it enables the use of a
\emph{single} SAT solver instance to run incremental BMC and ultimately the
entire \kiki algorithm.


A disadvantage of the approach is that many optimisations that can be
typically done during symbolic execution in order to simplify the
formula, such as constant propagation, can only be done locally, if at
all.



\section{The $\kiki$ algorithm} \label{sec:kiki}

As we outlined in Section~\ref{sec:approach}, the core of 2LS is the $\kiki$
algorithm, which we present in detail in this section. The general workflow of
the algorithm is shown in Figure~\ref{fig:kiki}.
Initially, $k = 1$ and $\tdom$ is
a set of predicates that can be used as invariant with $\top \in
\tdom$ (see Section~\ref{sec:inference} for details of how this is implemented).

After an initial test to see if any start states are
errors, $\kiki$ computes a $k$-inductive
invariant that covers the initial state and includes the assumption
that there are no errors in earlier states.
The invariant is then checked to see whether it is 
sufficient to show safety.  If there are possible reachable error
states then a second check is needed to see if the error is reachable
in $k$ steps (a genuine counterexample) or whether it is a potential
artefact of a too weak invariant.  In the latter case, $k$ is incremented
so that a stronger ($k$-)invariant can be found and the algorithm loops.

Also displayed in Figure~\ref{fig:kiki} are the steps of
incremental BMC, $k$-induction and classical over-approximating abstract
interpretation, given, respectively by the red dotted, blue dashed, and
green dashed/dotted boxes and arrows.
$\kiki$ can simulate $k$-induction by having
$\tdom = \{ \top \}$ and incremental BMC by over-approximating
the first SAT check.  Classical over-approximate abstract
interpretation can be simulated by having $\tdom = \adom$ for an abstract
domain $\adom$
and terminating with the result ``unknown'' if the first SAT check
finds a model.

\begin{figure}[t]

\begin{center}
\begin{tikzpicture}

\tikzset{stage/.style = {drop shadow, draw, fill=white, shape=rectangle}}
\tikzset{final/.style = {stage}}
\tikzset{merge/.style = {fill=black, shape=circle}}

\tikzset{result/.style = {}}


\tikzset{ibmc/.style = {red, densely dotted}}
\tikzset{kind/.style = {blue, dashed}}
\tikzset{absint/.style = {green, densely dashdotted}}

\tikzset{blink/.style = {thick}}
\tikzset{hlink/.style = {blink, -stealth, shorten >=3pt}}
\tikzset{tlink/.style = {blink, shorten <=3pt}}
\tikzset{link/.style = {hlink, blink, tlink}}

\tikzset{abox/.style = {ultra thin, rounded corners, fill, fill opacity=0.05}}

\path
  let
   \n{answerline} = {2cm},
   \n{outsideoffset} = {2.75cm},
   \n{rectx} = {6.5cm},
   \n{recty} = {3cm}
  in
   coordinate (answerline) at (0,-1*\n{answerline})
   coordinate (outsidelow) at (-1*\n{outsideoffset}, -1*\n{answerline})
   coordinate (start) at (0, 2*\n{recty})
   coordinate (init) at (0, 1.5*\n{recty})
   coordinate (outsidehigh) at (outsidelow |- init)
   coordinate (initsplit) at (0, \n{recty})
   coordinate (inv) at (\n{rectx}, \n{recty})
   coordinate (invce) at (\n{rectx}, 0)
   coordinate (invcesplit) at (0.5*\n{rectx}, 0)
   coordinate (concretece) at (0,0)
   coordinate (increment) at (0,0.6*\n{recty})
   coordinate (ce) at (answerline -| concretece)
   coordinate (unknown) at (answerline -| invcesplit)
   coordinate (safe) at (answerline -| invce)
 ;

\path
  let
   \n{step} = {0.75cm}
  in
   (start)
   +( 0, 0.5*\n{step}) coordinate (boxtop)
   +( 0,-0.5*\n{step}) coordinate (boxtoplower)
   +( 0,  \n{step}) coordinate (kikistart)
   +(-0.5*\n{step},0) coordinate (ibmcstartpart)
   +(-1*\n{step},0) coordinate (ibmcstart)
   +( 0.5*\n{step},0) coordinate (kindstartpart)
   +( 1*\n{step},0) coordinate (kindstart)
   coordinate (absintstart) at (kikistart -| inv)
  ;


\path
  let
   \n{insetwidth} = {0.6cm},
   \n{insetheight} = {1cm}
  in
 (ce)
 +(0,-1) coordinate (boxbottom)
 (invce)
 +(3,0) coordinate (boxright)
 (concretece)
 +(-3,0) coordinate (boxleft)
 (unknown)
 +(-0.5,0) coordinate (innerleft)
 +(-1*\n{insetwidth},0) coordinate (insetleft)
 +( 1*\n{insetwidth},0) coordinate (insetright)
 +( 0, 1*\n{insetheight}) coordinate (insettop)
 (concretece)
 +(2.7,0) coordinate (innerright)
 ;

\draw [abox, ibmc] (boxleft |- boxtop) -- (ibmcstartpart |- boxtop) -- (kindstartpart |- boxtoplower) -- (boxtoplower -| innerright) -- (innerright |- boxbottom) -- (boxleft |- boxbottom) -- cycle;

\node [anchor=north west, ibmc] at (boxleft |- boxtop) {IBMC};

\draw [abox, kind] (innerright |- boxtop) -- (kindstartpart |- boxtop) -- (ibmcstartpart |- boxtoplower) -- (boxtoplower -| boxleft) -- (boxleft |- boxbottom) -- (insetleft |- boxbottom) -- (insetleft |- insettop) -- (insettop -| insetright) -- (insetright |- boxbottom) -- (boxbottom -| boxright) -- (boxright |- increment) -- (innerright |- increment) -- cycle;

\node [anchor=north east, kind] at (innerright |- boxtop) {$k$-induction};

\draw [abox, absint] (boxtop -| innerleft) -- (innerleft |- boxbottom) -- (boxright |- boxbottom) -- (boxright |- boxtop) -- cycle;

\node [anchor=north east, absint, rectangle split, rectangle split parts=2, rectangle split draw splits=false] at (boxright |- boxtop) {Abstract \nodepart{second} Interpretation};

\node [merge] (Start) at (start) {};
\node [stage, rectangle split, rectangle split parts=2, rectangle split draw splits=false] (Init) at (init) {Test $\exists \vec{x}_0 \st$ \nodepart{second} $\init(\vec{x}_0) \land \serr(\vec{x}_0)$};

\node [merge] (Initsplit) at (initsplit) {};

\node [stage, rectangle split, rectangle split parts=3, rectangle split draw splits=false] (Inv) at (inv) {Find $\skinv \in \tdom \st \forall \vec{x}_0 , \dots, \vec{x}_k \st$ \nodepart{second}
$(\init(\vec{x}_0) \land \sprop[k] \land \strans[k-1] \limplies \sinv[k]) \land$ \nodepart{third}
$(\sprop[k] \land \sinv[k] \land \strans[k] \limplies \skinv(\vec{x}_k))$};

\node [stage, rectangle split, rectangle split parts=2, rectangle split draw splits=false] (Invce) at (invce) 
{Test $\exists \vec{x}_o , \dots, \vec{x}_k \st$ \nodepart{second} $\sprop[k] \land \sinv[k+1] \land \strans[k] \land \serr(\vec{x}_k)$};
\node [merge] (Invcesplit) at (invcesplit) {};

\node [stage, rectangle split, rectangle split parts=2, rectangle split draw splits=false] (Concretece) at (concretece)
{Test $\exists \vec{x}_0 , \dots, \vec{x}_k \st \init(\vec{x}_0) \land$ \nodepart{second} $\sprop[k] \land \sinv[k+1] \land \strans[k] \land \serr(\vec{x}_k)$};
\node [stage] (Increment) at (increment) {$k++$};

\node [final, starburst] (CE) at (ce) {\Large C/E};
\node [final, cloud, absint, fill=white] (Unknown) at (unknown) {\Large \color{black} ?};
\node [final, circle] (Safe) at (safe) {\Large Safe};


\draw [link] (kikistart) -- (Start.north);

\draw [link] (Start.south) -- (Init.north);
\node [result, right] (k0) at ($(Start.south)!0.5!(Init.north)$) {$k:=1$};
\draw [link] (Init.south) -- (Initsplit.north);
\node [result, right] (UNSAT) at ($(Init.south)!0.5!(Initsplit.north)$) {\result{UNSAT}};

\draw [link] (Initsplit.east) -- (Inv.west);

\draw [link] (Inv.south) -- (Invce.north);
\draw [link] (Invce.west) -- (Invcesplit.east);
\node [result, below] at ($(Invce.west)!0.5!(Invcesplit.east)$) {\result{SAT}};

\draw [link] (Invcesplit.west) -- (Concretece.east);

\draw [link] (Concretece.north) -- (Increment.south);
\node [result, left] at ($(Concretece.north)!0.5!(Increment.south)$) {\result{UNSAT}};
\draw [link] (Increment.north) -- (Initsplit.south);

\draw [link] (Invce.south) -- (Safe.north);
\node [result, left] at ($(Invce.south)!0.5!(Safe.north)$) {\result{UNSAT}};
\draw [link] (Concretece.south) -- (CE.north);
\node [result, left] at ($(Concretece.south)!0.5!(CE.north)$) {\result{SAT}};

\draw [tlink] (Init.west) -- (outsidehigh);
\draw [blink] (outsidehigh) -- (outsidelow);
\draw [hlink] (outsidelow) -- (CE.west);
\node [result, right] at (outsidehigh |- UNSAT) {\result{SAT}};

\draw [ibmc, link] (ibmcstart) -- (Start);
\draw [ibmc, link] (Initsplit.-60) -- (Concretece.17);

\draw [kind, link] (kindstart) -- (Start);
\draw [kind, link] (Initsplit.-30) -- (Invce.north west);

\draw [absint, link] (absintstart) -- (Inv.north);
\draw [absint, link] (Invcesplit.south) -- (Unknown);

\end{tikzpicture}
\end{center}
  
  \caption{The \kiki\ algorithm.}
  \label{fig:kiki}
\end{figure}



\subsection{Template-Based Predicate Inference} \label{sec:inference}


A key phase of \kiki\ is the generation of $\skinv$, a k-inductive invariant.
Perhaps the most obvious approach is to use an off-the-shelf abstract
interpreter.  
This works but will fail to exploit the real power of \kiki.
In each iteration, \kiki\ unrolls loops one more step (which can improve the
invariant given by an abstract interpreter) and adds assumptions that previous
unwindings do not give errors.

When directly using a solver, we would need to handle (the
existential fragment of) second-order logic.
As such solvers are not currently available, we reduce to a 
problem that can be solved by iterative application of a first-order
solver.
We restrict ourselves to finding \emph{k-inductive} invariants $\kinv$ of the form
$\templ(\vec{x},\vec{\delta})$ where $\templ$ is a fixed expression, a
so-called \emph{template}, over program variables $\vec{x}$ and
template parameters $\vec{\delta}$.
%
%
%
%
%
\begin{equation}\label{equ:invtempl}
\begin{array}{rrl}
\exists \vec{\delta} \st  & \forall \vec{x}_0 \dots \vec{x}_k \st & 
    \left(\init(\vec{x}_0) \land \strans[k-1] \limplies 
          \templ[k](\vec{\delta}) \right)
    \land \\
   &                 & 
    \left(\templ[k](\vec{\delta})
          \land \strans[k] \limplies \templ(\vec{x}_k,\vec{\delta}) \right)
\end{array}
\end{equation}
where $\strans[k]$ is the $k$-th unwinding of the transition relation and
$\templ[k]$ is a template for all states along the unwinding except for the
last state $\vx_k$:
\begin{eqnarray}
  \strans[k] = \bigland{i \in [0,k-1]}\trans(\vec{x}_i,\vec{x}_{i+1})
    \label{eq:strans}\\
  \templ[k](\vec{\delta}) = \bigland{i \in [0,k-1]} 
    \templ(\vec{x}_i,\vec{\delta}).
\end{eqnarray}
We resolve the
$\exists\forall$ problem by an iterative solving of the negated formula,
particularly of the second conjunct of \eqref{equ:invtempl}, for different
choices of constants $\vd$ as the values of the parameter $\vdelta$:
%
%
\begin{equation}\label{equ:qfinvtempl}
\exists \vec{x}_0 \dots \vec{x}_k \st \neg \big(\templ[k](\vec{d}) \wedge
\strans[k] \limplies \templ(\vec{x}_k,\vec{d})\big).
\end{equation}

The resulting formula can be expressed in quantifier-free logics and
efficiently solved by SMT solvers. Using this as a building block,
one can solve the mentioned $\exists\forall$ problem.

From the abstract interpretation point of view, $\vd$ is an abstract value,
i.e.\ it represents (\emph{concretises to}) the set of all program states
$\vs$---here, a state is a vector of values of variables from
$\vx$---that satisfy the formula $\templ(\vs, \vd)$. The abstract
values representing the infimum $\perp$ and supremum $\top$ of the abstract
domain denote the empty set and the whole state space, respectively:
\mbox{$\templ(\vs, \perp)$} $\equiv \false$ and $\templ(\vs, \top) \equiv
\true$~\cite{BJKS15}.  

Formally, the concretisation function $\gamma$ is:
$\gamma(\vd) = \{ \vs \mid \templ(\vs, \vd) \equiv \true \}.$
%
In the abstraction function, to get the most precise abstract value representing 
the given concrete program state $\vs$,
we let
$\alpha(\vs) = \{ \min(\vd) \mid \templ(\vs, \vd) \equiv \true\}$.
%
If the abstract domain forms a complete lattice, existence of such
a minimal value $\vd$ is guaranteed.

The algorithm for the invariant inference takes an initial value of $\vd =
\;\perp$ and iteratively solves~\eqref{equ:qfinvtempl} using an SMT solver.
If the formula is unsatisfiable, then an invariant has been found,
otherwise a model of satisfiability $\vd'$ is returned by the solver. The
model represents a counterexample to the current instantiation of the
template being an invariant. The value of the template parameter $\vd$
is then updated by combining the current value with the obtained model 
of satisfiability using a domain-specific join operator~\cite{BJKS15}.


For example, assume we have a program with a loop that counts from 0
to~10 in variable~$x$, and we have a template $x\leq d$. Let us assume
that the current value of the parameter~$d$ is~$3$, and we get a new
model $d'=4$.  Then we update the parameter to~$4$ by computing~$d
\sqcup d' = \max(d, d')$, because $\max$ is the join operator for
a~domain that tracks numerical upper bounds.

In 2LS, we use a single template to compute all invariants of the analysed
program.
Therefore, typically, a template is composed of multiple parts, each part
describing an invariant for a set of program variables.
With respect to this, we expect a template $\templ(\vx, \vdelta)$ to be composed
of so-called \emph{template rows} $\templ_r(\vx_r, \delta_r)$, each row $r$
describing an invariant for a subset $\vx_r$ of variables $\vx$ and having its
own row parameter $\delta_r$.
The overall invariant is then a composition of individual template rows with
computed values of the corresponding row parameters.
The kind of the composition (it can be, e.g., a simple conjunction) is defined
by each domain.

\subsubsection{Guarded Templates} Since we use the SSA form rather than control
flow graphs, we cannot use templates directly.
Instead we use \emph{guarded templates}.
As described above, a template is composed of multiple template rows, each row
describing an invariant for a subset of program variables.
In a guarded template, each row $r$ is of the form $G_r(\vx_r) \limplies
\basetempl_r(\vx_r,\delta_r)$ for the $r^{\text{th}}$ row $\basetempl_r$ of the
base template domain (e.g.  template polyhedra).
$G_r$ is the conjunction of the SSA guards $g_r$ associated with the definition
of variables $\vx_r$ occurring in $\basetempl_r$.
Since we intend to infer loop invariants, $G_r(\vx_r)$ denotes the guard
associated to variables $\vx_r$ appearing at the loop head. 
Hence, template rows for different loops have different guards.

%

We illustrate the above on the example program in Figure~\ref{fig:ssa_loops}
using a guarded interval template.
The template has the form: \begin{equation}\label{eq:ex_templ}
\templ(\ssaxlb{6},(\delta_1,\delta_2)) = \begin{array}{lcr} \ssag{3} \wedge
\ssagls{6} &\limplies & \ssaxlb{6} \leq \delta_1 \;\wedge\\[0.3em] \ssag{3}
\wedge \ssagls{6} &\limplies & -\ssaxlb{6} \leq \delta_2.  \end{array}
\end{equation}
Here, $\ssag{3}$ and $\ssagls{6}$ guard the definition of
$\ssaxlb{6}$---$\ssag{3}$ expresses the fact that the loop head is reachable and
$\ssagls{6}$ expresses that $\ssaxlb{6}$ is chosen as the value of
$\ssaphi{x}{3}$.

\subsubsection{Solving of the $\exists\forall$ Problem}
\label{sec:qe}

As discussed above, it is necessary to solve an $\exists\forall$ problem to
find values for template parameters $\vec{\delta}$ to infer invariants.

\subsubsection*{Model enumeration}
The well-known method \cite{RSY04,BKK11} for solving this problem
in~\eqref{equ:qfinvtempl} using an SMT solver is to repeatedly check
satisfiability of the formula for different abstract values $\vec{d}$ (starting
with the infimum $\vec{d}=\bot$):
\begin{equation}\label{equ:indcheck}
  \templ[k](\vec{d}) \land \strans[k] \wedge \neg
  \templ(\vec{x}_k,\vec{d}).
\end{equation}

If it is unsatisfiable, then we have found an invariant; otherwise, we join the
model returned by the solver with the previous abstract value $\vec{d}$ and
repeat the process with the new value of $\vd$ obtained from the join.

This method corresponds to performing a classical Kleene iteration on the
abstract lattice up to convergence.
Convergence is guaranteed because our abstract domains are finite. 
However, while this method might be sufficient for some abstract domains
(especially those with a low number of possible states), it is practically
unusable for other ones.
For example, when dealing with integer variables, the height of the lattice is
enormous and even for a one-loop program incrementing an unconstrained 64-bit
variable, the na\"ive algorithm will not terminate within human life time.
Hence, for some abstract domains, we are going to use an optimised method
(e.g., for some numerical domains, see Section~\ref{sec:tempoly}).

In general, there may be a specific method for solving the $\exists\forall$
problem for each domain.
We refer to these methods as to \emph{domain strategy iterations}.

\subsubsection*{Generic domain strategy iteration}

Even though each domain may implement its own strategy iteration algorithm, we
observe that all algorithms are to some extent similar.
This is related to the fact that abstract domain templates are typically
composed of multiple template rows, as described earlier in this section.

With respect to this, we developed a \emph{generic strategy iteration}
algorithm~\cite{Marusak19} parametrised by an abstract domain having the form
of a template.
The algorithm is shown in Figure~\ref{alg:generic}.

\begin{figure}[h!]
\begin{algorithmic}[1]
  \STATE $\vd \gets \perp$
  \WHILE {$\templ[k](\vd)$ is not an invariant}
  \STATE solver $\gets \templ[k](\vd)$
    \STATE solver $\gets \neg \templ(\vx_k, \vd)$
    \IF {solver.solve() $=$ SAT}
      \FOR {$\templ_r(\vx_r, d_r) \in \templ(\vx, \vd)$ }
        \STATE \changed{$m_r \gets$ solver.model($\vx_r$)} 
        \STATE \changed{find $d_m$ s.t. $\templ_r(m_r, d_m)$ holds}
        \STATE \changed{$d_r \gets d_r \sqcup d_m$}
      \ENDFOR
    \ENDIF
  \ENDWHILE
\end{algorithmic}
\caption{Generic strategy iteration algorithm for solving the $\exists\forall$
problem.}
\label{alg:generic}
\end{figure}

The algorithm repeatedly solves~\eqref{equ:indcheck} for the given
abstract domain. 
If the formula is satisfiable, then for each template row $r$, the algorithm
gets the model of satisfiability for the variables that $r$ describes.
\changed{
The obtained model (i.e., the values of the concerned variables) is used to
instantiate the template row formula (line 8) and the corresponding value of
the template parameter is joined with the previous value of the row parameter.
}

Since 2LS uses incremental solving, we assume that the transition relation
(expressed by the SSA form) is already a part of the solver clause set.
Moreover, formulae added to the solver clause set on lines 3 and 4 are removed
after each iteration.
This ensures efficiency of the method since only the new formulae need to be
re-solved every time.

However, as we mentioned in the previous section, an optimisation may be
required in order to assure scaling of the algorithm.
In such case, lines 7-9 are replaced by the optimised method for
determining values of template row parameters.
We will give an example for such optimised techniques in
Section~\ref{sec:tempoly}.

\subsection{Incremental Bounded Model Checking}\label{sec:incrbmc}


Bounded Model Checking (BMC)~\cite{BCCZ99}  
focuses on refutation by picking an
\define{unwinding limit} $k$ and solving the equation
\begin{equation}\label{eq:bmc}
  \exists \vec{x}_0, \dots, \vec{x}_k \st \init(\vec{x}_0) \land
  \strans[k]
  \land
  \lnot \sprop[k+1]
\end{equation}
where $\strans[k]$ is an unwound transition relation as defined
by~\eqref{eq:strans} and $\sprop[k]$ is a predicate stating that $k$ states are
error-free:
\begin{equation}
  \sprop[k]  = \bigland{i \in [0,k-1]} \lnot \err(\vec{x}_i).
\end{equation}
Models of~\eqref{eq:bmc} correspond to concrete
counterexamples of some length $n \lte k$.  The unwinding limit gives
an \emph{under-approximation} of the set of reachable states and thus
can fail to find counterexamples that take a large number of
transition steps.  In practice, BMC works well as the formula is
existentially quantified and thus is in a fragment handled well by SAT
and SMT solvers.

Incremental bounded model checking is one of the core components
of the \kiki algorithm presented in Section~\ref{sec:kiki}.
It corresponds to the red part in Figure~\ref{fig:kiki}.
Incremental BMC (IBMC)~(e.g.~\cite{ES03b}) uses repeated BMC checks (often
optimised by using the solver incrementally) with increasing
bounds to avoid the need for a fixed bound.
If the bound starts at 0
(i.e.~checking $\exists x_0 \st \init(\vec{x}_0) \land
\err(\vec{x}_0)$) and is increased by one in each step (this is the common
use-case), then it can be assumed that there are no errors at previous
states, giving a simpler test:

\begin{equation}
  \exists \vec{x}_0, \dots, \vec{x}_k \st 
  \init(\vec{x}_0) \land 
  \strans[k] \land \sprop[k]
  \land
  \err(\vec{x}_k).
\end{equation}

In Section~\ref{sec:sat}, we will discuss how incremental BMC
can be implemented efficiently using incremental SAT solving.

\subsection{Incremental $k$-Induction}\label{sec:kinduction}


Incremental $k$-induction~\cite{SSS00} is the blue part of the \kiki algorithm
in Figure~\ref{fig:kiki}.
It can be viewed as an extension of IBMC 
that can show system
safety as well as produce counterexamples.  It makes use of
\emph{$k$-inductive invariants}, which are predicates that have the
following property:
\begin{equation}
  \forall \vec{x}_0 \dots \vec{x}_k \st \sinv[k] \land \strans[k] \limplies \kinv(\vec{x}_k)
\end{equation}
\noindent where
\begin{equation*}
  \sinv[k] = \bigland{i \in [0,k-1]} \kinv(\vec{x}_i).
\end{equation*}
\noindent $k$-inductive invariants have the following useful properties:
\begin{itemize}
\item{Any inductive invariant is a $1$-inductive invariant and vice versa.}
\item{Any $k$-inductive invariant is a $(k+1)$-inductive invariant.}
\item{A (finite) system is safe if and only if there is a $k$-inductive
  invariant $\kinv$ which satisfies:
\begin{equation}
\begin{array}{rrl}
   & \forall \vec{x}_0 \dots \vec{x}_k \st & 
    \left(\init(\vec{x}_0) \land \strans[k] \limplies \sinv[k] \right)
    \land \\
   &                 & 
    \left(\sinv[k] \land \strans[k] \limplies \kinv(\vec{x}_k)\right)
    \land \\
   &                 & 
    \left(\kinv(\vec{x}_k) \limplies \lnot \err(\vec{x}_k)\right).
\end{array}
\end{equation}
}
\end{itemize}
\noindent Showing that a $k$-inductive invariant exists is sufficient to
show that an inductive invariant exists \emph{but it does not imply
  that the $k$-inductive invariant is an inductive invariant}.  Often
the corresponding inductive invariant is significantly more complex.
Thus $k$-induction can be seen as a trade-off between 
invariant \emph{generation} and \emph{checking}
as it is a means to benefit as much as possible from 
simpler invariants by using a more complex property check.

However, finding a candidate $k$-inductive invariant is still hard, and so
implementations often use $\lnot \err(\vec{x})$ as the candidate. Similarly to
IBMC, linearly increasing $k$ can be used to simplify the expression by
assuming there are no errors at previous states:
\begin{equation}
\begin{array}{rrl}
 & \exists \vec{x}_0, \dots, \vec{x}_k \st &
  (\init(\vec{x}_0) \land \strans[k] \land \sprop[k] \land \err(\vec{x}_k)) \lor\\
 &                 & 
  (\strans[k] \land \sprop[k] \land \err(\vec{x}_k)).
\end{array}
\end{equation}
\noindent A model of the first part of the disjunct is a concrete
counterexample ($k$-induction subsumes IBMC) and if the whole
formula has no models, then $\lnot \err(\vec{x})$ is a $k$-inductive
invariant and the system is safe.

\subsection{Incremental SAT Solving} \label{sec:sat}

2LS requires incremental back-end solvers.
%
%
%
%
%
The first ideas for incremental SAT solving date back to the 1990s~
\cite{Hoo93,SS97,KWSS00}.  The question is how to solve a sequence
of similar SAT problems while reusing the effort spent on solving previous
instances.  The authors of~\cite{Str01,WKS01} identify conditions for the
reuse of learnt clauses, but this requires expensive book-keeping, which
partially saps the benefit of incrementality.
Obviously, incremental SAT solving is easy when the modification to
the CNF representation of the problem makes it grow monotonically.
This means that if we want to solve a sequence of (increasingly
constrained) SAT problems with CNF formulae $\Phi(k)$ for $k\geq 0$,
then $\Phi(k)$ must be \emph{growing monotonically} in $k$, i.e.,
$\Phi(k+1) = \Phi(k) \wedge \varphi(k)$ for CNF formulae $\varphi(k)$.
Removal of clauses from $\Phi(k)$ is trickier as some of the clauses
learnt during the solving process are no longer implied by the new
instance and need to be removed as well.  
This requires additional solver features like solving \emph{under
  assumptions}~\cite{ES03b}, which is the most popular approach to
incremental SAT solving: assumptions are temporary assignments to variables 
that hold solely for one specific invocation of the SAT solver.
We will see that incremental BMC requires 
a \emph{non-monotonic} series of formulae.

An alternative approach is to use SMT solvers. 
SMT solvers offer an interface for pushing and popping clauses in a stack-like
manner. Pushing adds clauses, popping removes them from the formula. 
This makes the modification of the formula intuitive to the user, but the
efficiency depends on the underlying implementation of the push and pop
operations.
For example, in \cite{GW14}, it was observed that some SMT solvers (like Z3) are
not optimised for incremental usage and hence perform worse incrementally than
non-incrementally.

Consequently, since the support for incremental solving in SMT solvers is still
lagging behind in comparison to SAT solvers, we use SAT solving.
The CPROVER framework~\cite{CKL04} itself implements powerful
bitvector decision procedures that use a SAT solver such as
\minisat~\cite{ES03a} or Glucose%
\footnote{\url{http://www.labri.fr/perso/lsimon/glucose/#glucose-4.0}}
as the backend solver.  For SAT solvers, solving under assumptions is the
prevalent method.

\subsubsection*{Formula construction in incremental BMC}
%
%
Following the construction in \cite{ES03b} (stated for finite state machines),
incremental BMC can be formulated as a sequence of SAT problems
$\Phi(k)$ that we need to solve:
\begin{equation}\label{equ:incrbmc}
\begin{array}{lcl}
\Phi(0) &:=& \init(\vec{x}_0) \wedge (\alpha_0 \Rightarrow \err(\vec{x}_0))\\
                 && \text{with the assumption }\alpha_0,\\
\Phi(k+1) &:=& \Phi(k) \wedge \neg \err(\vec{x}_k) \wedge \trans(\vec{x}_k,\vec{x}_{k+1})  \wedge \neg \alpha_k
 \wedge (\alpha_{k+1} \Rightarrow \err(\vec{x}_{k+1}))\\ 
                 && \text{with the assumption } \alpha_{k+1}\\
\end{array}
\end{equation}
where $\alpha_k$ are Boolean variables.
Let us now take a look why these are necessary.
Consider a program counting in variable $x$ from 0 to 10.
There is an assertion inside the loop that $x\leq 1$ is expected to hold. We
start with the initial formula (omitting Boolean variables $\alpha_k$):
$$
\Phi(0) := (x_0{=}0) \wedge (x_0{<}10)\wedge (x_1{=}x_0+1)\wedge \neg(x_1{\leq}1).
$$
This formula is UNSAT. We now (na\"ively) extend the formula with the next
loop iteration:
$$
\Phi(1) := (x_0{=}0) \wedge (x_0{<}10)\wedge (x_1{=}x_0{+}1)\wedge \neg(x_1{\leq}1)
\wedge (x_1{<}10)\wedge (x_2{=}x_1{+}1)\wedge \neg(x_2{\leq}1).
$$
This is still UNSAT, although we expect it to be SAT. Obviously, the reason is
that we are adding further conjuncts to something that is already UNSAT.
Therefore, in order to be sound, we have to remove $x_1>1$ from the previous
iteration, which makes the formula construction non-monotonic.

However, as explained above, we cannot
simply remove parts of the formula from the solver, but 
we can solve this issue with the help of \emph{solving under assumptions}.
In iteration $k$, the assumption $\alpha_k$ is assumed to be true, whereas
it is assumed false for iterations $k'>k$. This has the effect that, 
in iteration $k'$, the sub-formula $(\alpha_k \Rightarrow \err(\vx_k))$ 
becomes trivially satisfied. Hence, it does not contribute to the
(un)satisfiability of $\Phi(k')$, which emulates its deletion.%
\footnote{For a large number of iterations $k$, such
  trivially satisfied subformulas might accumulate as ``garbage'' in the
  formula and slow down its resolution. Restarting the solver 
  at appropriate moments is the common solution to this issue~\cite{DBLP:journals/fmsd/CabodiCMPP17}.}

\subsubsection*{The Benefit of Incremental BMC}
%
\begin{figure*}[t]
\centering
\hspace*{-10em}\includegraphics[scale=0.68]{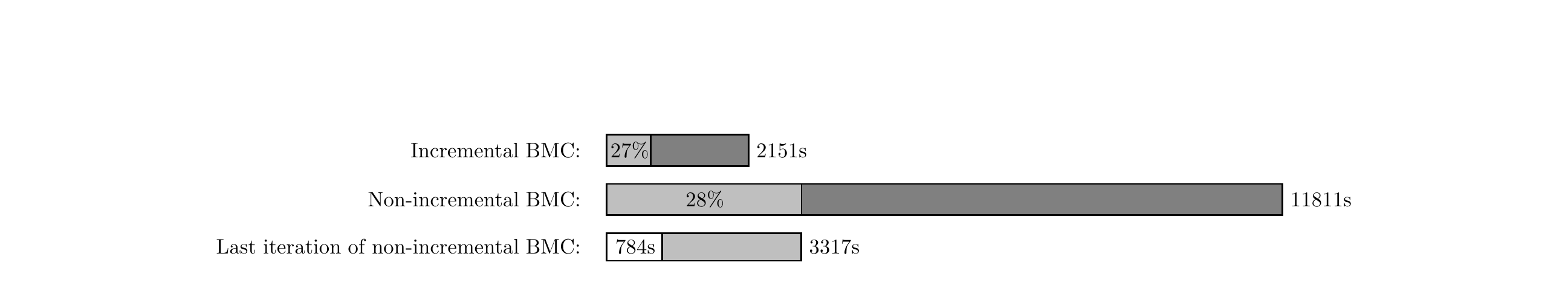}
\caption{\label{fig:solving}
  The benefit of incremental BMC: solving time vs. overall runtime
  (explanation in the text).}
\end{figure*}

Non-incremental BMC has to perform extra work in symbolic execution
for each unwinding $k$.
As investigated in~\cite{SKB+17},
one might argue that removing this overhead is the main reason why
incremental BMC is faster.
However, the overhead for symbolic execution when compared to generating and
solving a SAT formula is similar for the incremental and non-incremental
approach:
on the embedded software benchmark set of \cite{SKB+17},
27\,\% of the time taken by the incremental approach are spent in
solving SAT formulae (582 out of 2,151 seconds), compared with 28\,\%
of the time taken by the non-incremental approach (3,317 out of 11,811
seconds).
We~illustrate this observation in the bar chart in
Fig.~\ref{fig:solving}, which plots the total runtime consisting of
the time spent in generating SAT formulae and solving them (light
grey) and the overhead (dark grey) for incremental and non-incremental
BMC.
Unsurprisingly, as shown in the third bar in Fig.~\ref{fig:solving},
solving the instance for the largest $k$ in the non-incremental
approach (white) takes a considerable amount of time (around 24\,\%),
when compared to the total time (white+grey) for solving the SAT
formulae for iterations 1 to $k$ (784 out of 3,317 seconds).

An explanation for the speedups achieved by incremental BMC might be
the size of the queries issued in both approaches.  The average number
of clauses per solver call is halved from 1,367k clauses for the
non-incremental approach to 709k clauses for the incremental
approach. Similarly, the average number of variables of the last call
is less than a third in the incremental approach when compared to the
non-incremental approach, being 217k and 746k respectively.  That
means that the SAT solver is able to simplify the formulae within
pre-processing much more efficiently in the incremental approach.

The work~\cite{SKB+17} experimented with an incremental slicing algorithm in
combination with incremental unwinding in \cbmc; however, this is not
implemented \twols.
The work~\cite{MSH+17} developed an SMT solving algorithm in 2LS (which
is available in a prototype branch of 2LS) that uses template polyhedra
instead of Boolean literals in order to perform theory-level propagation based
on the concept of Abstract Conflict Driven Learning~\cite{DHK13}.

\section{Implemented Analyses} \label{sec:features}


2LS supports analysis of various program features such as reachability of
assertions, termination, or memory safety. For most of these analyses, 2LS
introduces abstract domains~\cite{CC77} for invariant inference, which is one of the steps
of the \kiki algorithm.

There are numerical domains such as template polyhedra (see
Section~\ref{sec:tempoly}), equalities and disequalities, domains for
ranking functions (see Section~\ref{sec:termination}) and recurrent
sets(see Section~\ref{sec:nontermination}), and domains for the shape of
data structures (see Section~\ref{sec:memorysafety}).

In this section, we introduce the most important supported abstract domains as
well as other concepts that support verification of various program properties.

\subsection{Template Polyhedra Abstract Domain}\label{sec:tempoly}


Template polyhedra \cite{SSM05} are
a class of templates for numerical variables which have the form
$\templ = (\mathbf{A}\vec{x}\leq\vec{\delta})$ where
$\mathbf{A}$ is a matrix with fixed coefficients.
Subclasses of such templates include \emph{Intervals},
which require constraints $\vecv{1}{-1}x_i\leq\vecv{\delta_{i1}}{\delta_{i2}}$
for each variable $x_i$, \emph{Zones} (differences), and
\emph{Octagons} \cite{Min01a}.
The $r^{\textit{th}}$ \emph{row} of the template are the constraints
generated by the $r^{\text{th}}$ row of matrix $\mathbf{A}$.

In our template expressions, variables $\vec{x}$ are
\emph{bit-vectors} representing signed or unsigned integers. 
These variables can be mixed in template constraints. Type promotion
rules are applied such that the bit-width of the types of the
expressions are extended in order to avoid arithmetic under- and
overflows in the template expressions.
$\top$ corresponds to the respective maximum values in the promoted
type, whereas $\bot$ must be encoded as a special symbol.

\subsubsection*{Optimised Solving of the $\exists\forall$ Problem}

In order to solve the parameter synthesis problem efficiently, we need a
convergence acceleration that makes the computational effort \emph{independent}
from the number of states and loop iterations.
To this end, we use a technique that is inspired by an encoding used by
max-\emph{strategy iteration} methods \cite{GS07,GM11,MS14b}.
These methods state the invariant inference problem over template polyhedra as a
disjunctive linear optimisation problem, which is solved iteratively by an
upward iteration in the lattice of template polyhedra:
using SMT solving, a conjunctive subsystem (``strategy'') whose
solution extends the current invariant candidate is selected.  This
subsystem is then solved by an LP solver; the procedure terminates as
soon as an inductive invariant is found.\footnote{
  Intuitively, this works as follows: As an example,
  assume the domain of upper bounds on numerical variables (but the
  approach extends to any domain satisfying the conditions mentioned
  further below). An invariant upper bound at a program location is
  then the maximum value contributed by any of the incoming
  transitions to the program location. Writing this down as an
  optimisation problem over a system of linear inequalities, incoming
  transitions form disjunctions. Considering multiple program
  locations we get conjunctions of these disjunctions. So, we have a
  linear system in the CNF form. This can be solved by finding those
  disjuncts (i.e., incoming transitions) in each clause that are
  satisfied by the initial state (there must be such, otherwise the
  system is not feasible). We can then pick these disjuncts and
  conjoin them (as the ``strategy''), resulting in a (conjunctive)
  linear system (called a conjunctive subsystem), which can be solved
  using an LP solver. The obtained upper bound is only valid for those
  disjuncts that have been picked. If it is not an invariant yet, then
  there must be disjuncts that can be substituted for some of the
  disjuncts to improve the bound. This can be repeated and is
  guaranteed to terminate with the minimal invariant.}

The above method can only be used if the domain is convex and the parameter
values are ordered and monotonic w.r.t. concretisation, which holds true, for
example, for template polyhedra $\mathbf{A}\vec{x}\leq\vec{d}$ where $\vec{d}$
is a parameter but not for those where $\mathbf{A}$ is a parameter.
If the operations in the transition relation satisfy certain properties such as
monotonicity of condition predicates, then the obtained result is the least
fixed point, i.e. the \emph{same} result as the one returned by the na\"ive
model enumeration above but much faster on average.

We adapt this method to our setting with bit-vector variables and
guarded templates. Since we deal with finite domains (bit-vectors), we
can use \emph{binary search} as an optimisation method instead of an LP solver.

The algorithm proceeds as follows:
We start by checking whether the current abstract value $\vec{d}$ (starting from
$\vec{d}=\bot$) is inductive (see~\eqref{equ:indcheck}). If so, we have found an
invariant; otherwise there are template rows $R$ whose values are not inductive
yet.
Moreover, the counterexample to induction obtained from the inductivity
check tells us which bounds need to be improved.
We construct the system 
\begin{equation}\label{equ:symbsys}
\bigland{i \in [0,k-1]} \left\{\begin{array}{rll}
& \bigwedge_{r\notin R} & G_r(\vec{x}_i) \limplies (\rowexpr_r(\vec{x}_i)\leq d_r) \\[1ex]
\wedge & \bigwedge_{r\in R} & G_r(\vec{x}_i) \limplies (\rowexpr_r(\vec{x}_i)\leq \delta_r)\\[1.5ex]
\end{array}\right\} 
\wedge \strans[k]
\wedge \bigwedge_{r\in R} (G_r(\vec{x}_k) \wedge \delta_r\leq \rowexpr_r(\vec{x}_k))
\end{equation}
where $\rowexpr_r$ is the left-hand side of the inequality
corresponding to the $r^\text{th}$ row of the template.
Then we start the binary search for the minimal value of $\sum_{r\in R}\delta_r$ 
over this system.
The initial bounds for $\sum_{r\in R}\delta_r$ are as follows:
\begin{itemize}
\item The lower bound $\ell$ is $\sum_{r\in R} d'_r$ where $d'_r$ is the
  value of $\rowexpr_r(\vec{x}_k)$ in the model of the inductivity
  check~\eqref{equ:indcheck} above;
\item The upper bound $u$ is $\sum_{r\in R}\mathit{max\_value}(r)$ where
  $\mathit{max\_value}$ returns the maximum value that $\rowexpr_r(\vec{x}_k)$
  may have (dependent on variable type).
\end{itemize}

The binary search is performed by iteratively checking~\eqref{equ:symbsys} for
satisfiability under the assumption
$\sum_{r\in R}\delta_r\geq m$ where $m = \mathit{median}(\ell,u)$.  If
satisfiable, set $\ell := m$, otherwise set $u := m$ and repeat until
$\ell = u$.
The values of $\delta_r$ in the last satisfiable query are assigned to 
$d_r$ to obtain the new abstract value.
The procedure is then repeated by testing whether $\vec{d}$ is
inductive~\eqref{equ:indcheck}.
Note that this algorithm uses a similar encoding for bound
optimisation as strategy iteration, but it potentially requires a higher
number of iterations than strategy iteration.  This choice has been
made deliberately in order to keep the size of the generated SMT
formulas small, at the cost of a potentially increased number of
iterations.
The same row $r$ may be removed from $R$ and improved again in
later iterations. However, as each iteration makes progress,
the algorithm terminates in a finite number of iterations.

We illustrate the binary search
algorithm on the example from Figure~\ref{fig:ssa_loops}. 
We use the interval abstract domain to compute the inverval for the value of the
variable $x$ at the end of the loop. 
We let $k = 1$, and therefore we let $\vx_0 = (\ssaxlb{6})$ and $\vx_1 =
(\ssax{5})$.\footnote{Variables $\vx_0$ apply to the case of zero
iterations of the loop, hence unconstrained $\ssaxlb{6}$ is taken here as the
value of $x$.}
The corresponding guarded template $\templ$ has the form defined
in~\eqref{eq:ex_templ}.
The template contains two template rows defined as: \begin{eqnarray} &
G_0(\ssaxlb{6}) \limplies \ssaxlb{6} \leq \delta_0 \\ & G_1(\ssaxlb{6})
\limplies -\ssaxlb{6} \leq \delta_1 \end{eqnarray} where $G_0(\ssaxlb{6}) =
G_1(\ssaxlb{6}) = \ssag{3} \wedge \ssagls{6}$.
We also define the template row guards for variables $\vx_1$ (corresponding to
the values of the program variables after one execution of the transition
relation) as $G'_0(\ssax{5}) = G'_1(\ssax{5}) = \ssag{5} = \ssaxphi{3} < 10$.

Initially, $\vd = \perp$, and we solve~\eqref{equ:indcheck}.
Since using $\vd = \perp$ as the value of the parameters makes the
instance of the template $\false$ by definition, the invariant cannot be
inductive, and we get a model of satisfiability.
Neither of the template rows is inductive, and hence we
instantiate~\eqref{equ:symbsys} with $R = \{0, 1\}$.
The obtained system of formulae is the following (with $\trans(\vx_0,
\vx_1)$ being the conjunction of the constraints shown in
Fig.~\ref{fig:ssa_loops}(b)):
\begin{equation}
\begin{split}
  G_0(\ssaxlb{6}) \limplies \ssaxlb{6} \leq \delta_0 ~\wedge \\
  G_1(\ssaxlb{6}) \limplies -\ssaxlb{6} \leq \delta_1 ~\wedge \\
  \trans(\vx_0, \vx_1) ~\wedge \\
  G'_0(\ssax{5}) \wedge \delta_0 \leq \ssax{5} ~\wedge \\
  G'_1(\ssax{5}) \wedge \delta_1 \leq -\ssax{5}. \\
\end{split}
\end{equation}
We apply binary search to the system starting with $\ell = -2^{31}$ and $u =
2^{31} - 1$.
We get $\delta_0 = 1$ and $\delta_1 = -1$.
Now, template row 1 ($-\ssaxlb{6} \leq -1$) is inductive, however,
template row 0 is not inductive.
We construct a new system as per~\eqref{equ:symbsys} with $R = \{0\}$:
\begin{equation}
\begin{split}
  G_0(\ssaxlb{6}) \limplies \ssaxlb{6} \leq \delta_0 ~\wedge \\
  G_1(\ssaxlb{6}) \limplies -\ssaxlb{6} \leq -1 ~\wedge \\
  \trans(\vx_0, \vx_1) ~\wedge \\
  G'_0(\ssax{5}) \wedge (\delta_0 \leq \ssax{5}).
\end{split}
\end{equation}
Using binary search, we get $\delta_0 = 10$.
With this, template row 0 is inductive, and so we have found an
invariant, namely:
\begin{equation}
\begin{split}
  & \ssag{3} \wedge \ssagls{6} \limplies \ssaxlb{6} \leq 10 ~\wedge \\
  & \ssag{3} \wedge \ssagls{6} \limplies -\ssaxlb{6} \leq -1. 
\end{split}
\end{equation}

\subsubsection*{Bitvector Width Extension}

Integers wrap around on most architectures when
they over/underflow. Thus, we have to be mindful of these in
arithmetic within templates in order to guarantee soundness of the results.

Let us consider an example of analysing termination of the following function
by synthesising a ranking function.
\begin{lstlisting}[numbers=none]
void f() {
  signed char x;
  while(1) x++;
}
\end{lstlisting}

The ranking function synthesis aims to compute a value for a template
parameter $\rankparam$ such that $\rankparam(x{-}x')>0$ holds for all $x,
x'$ under the transition relation $x'{=}x{+}1$ and the computed invariant $\true$.

\medskip
Thus, assuming that the current value for $\rankparam$ is $-1$, the
constraint to be solved is
\[ \true \wedge x'{=}x{+}1 \wedge \neg({-1}\cdot(x{-}x'){>}0), \]
which is equivalent to $\neg({-1}(x{-}(x{+}1)){>}0)$.
While for mathematical integers this is satisfiable, it is unsatisfiable for
signed bit-vectors due to overflows.  For $x{=}127$, the overflow happens
such that $x{+}1{=}{-}128$ if \texttt{signed char} is an 8-bit signed integer.
Thus, $\neg(-1 \cdot (127{-}({-}128)){>} 0)$ becomes $\neg (1 {>} 0)$,
which makes the constraint unsatisfiable, and we would incorrectly conclude
that $-x$ is a ranking function, which does not hold for signed bitvector
semantics.  However, if we extend the bitvector width to $k{=}9$ such that
the arithmetic in the template does not overflow, then
$\neg(-1\cdot((\mathit{signed}_9)127{-}(\mathit{signed}_9)({-}128))>0)$
evaluates to $\neg((-1 \cdot 255)>0)$, where $\mathit{signed}_k$ is a cast to a $k$-bit
signed integer.
Now, $x{=}127$ is a model that shows that $-x$ is not a valid ranking
function.

For these reasons to retain soundness, we extend the bit-width of
signed and unsigned integer operands to integers that can hold the
result of the operation without over- or underflow, e.g., one
additional bit for additions and doubling the size plus one bit for
multiplications.
The maximum bit-width required depends on the shape of the
template. Since our templates contain a finite number of operations,
the maximum bit-width is finite.

Floating-point numbers do not require extensions as they overflow
(resp. underflow) to infinity (resp. minus infinity), which does not
impact soundness.












\subsection{Termination Analysis}\label{sec:termination}


For reasoning about termination, we need the notions of ranking
functions and preconditions for termination.

\begin{definition}[Ranking function] \label{def:ranking_function}
A \emph{ranking function} for a procedure $(\init,$ $\trans)$
is a function $r$ such that
\[
  \begin{array}{rlr}
  \multicolumn{2}{l}{\exists \Delta>0, \inv: \forall \vx, \vxp:} \\
   & \left(\init(\vx) \Longrightarrow \inv(\vx)\right)
  \\
  \wedge & \left(\inv(\vx)\wedge\trans(\vx,\vxp)\Longrightarrow \inv(\vxp)
  \wedge r(\vx) {-} r(\vxp) {>} \Delta \wedge r(\vx) {>} 0\right).
  \end{array}
  \]
Thus, $r$ is a function from the set of program states to a well-founded domain, e.g. $\mathbb{R}^{\geq 0}$.
\end{definition}

We denote by $\rrank(\vx,\vxp)$ the constraints on $r$ that form the
\emph{termination argument}, i.e., $r(\vx) - r(\vxp) > \Delta \wedge r(\vx) >
0$ for monolithic ranking functions.
The existence of a ranking function for a procedure guarantees its
\emph{universal} termination.

Monolithic ranking functions are complete, i.e., termination can always be
proven monolithically if a program terminates.  However, in practice,
combinations of linear ranking functions, e.g., linear lexicographic
functions
are preferred.  This is driven by the fact that
monolithic \emph{linear} ranking functions are not expressive enough, and
that \emph{non-linear} theories are challenging for the existing SMT
solvers, which handle the linear case much more efficiently.

\begin{definition}[Lexicographic ranking function] 
  \label{def:lexicographic} A \emph{lexicographic ranking function~$R$}
  for a procedure $(\init, \trans)$ is an
  $n$-tuple of expressions $(\rank_n, \rank_{n-1}, \ldots,\allowbreak \rank_1)$ such
  that
$$
\begin{array}{rllr}
  \multicolumn{3}{l}{\exists \Delta>0, \inv: \forall \vx, \vxp:} \\
   & \multicolumn{2}{l}{\left(\init(\vx) \Longrightarrow \inv(\vx)\right)}\\
  \wedge & \multicolumn{2}{l}{\big(\inv(\vx)\wedge
    \trans(\vx,\vxp)\Longrightarrow \inv(\vxp)
  \wedge \exists i\in [1,n]:}\\
&& \rank_i(\vx) > 0 & \text{(Bounded)} \\
&~~~~\wedge & \rank_i(\vx) - \rank_i(\vxp) > \Delta & \text{(Decreasing)} \\
&~~~~\wedge & \forall j>i: \rank_j(\vx) - \rank_j(\vxp) \ge 0~\big) &
\qquad\text{(Non-increasing)}.
\end{array}
$$
\end{definition}
Notice that this is a special case of
Definition~\ref{def:ranking_function}.  In particular, the existence
of $\Delta>0$ and the \emph{Bounded} condition guarantee that $>$ is a
well-founded relation.

Before we encode the requirements for lexicographic ranking functions into
constraints, we need to optimise them to take advantage of bit-vector
semantics.  Since bit-vectors are bounded, it follows that the
\emph{Bounded} condition is trivially satisfied and therefore can be
omitted.  Moreover, bit-vectors are discrete, hence we can replace the
\emph{Decreasing} condition with $\rank_i(\vx)-\rank_i(\vxp)>0$.  The
following formula, $\lexrank^n$, holds if and only if $(\rank_n, \rank_{n-1},
\ldots, \rank_1)$ is a lexicographic ranking function with $n$ components over
bit-vectors.
\[
\lexrank^n(\vx,\vxp) = 
\bigvee_{i=1}^{n} \left(\rank_i(\vx)-\rank_i(\vxp)>0 \; \wedge
  \bigwedge_{j=i+1}^{n}(\rank_j(\vx)-\rank_j(\vxp)\geq 0)\right)
\]
Assume we are given the transition relation $\trans(\vx,\vxp)$ of a
procedure $f$.  The procedure $f$ may be composed of several
loops, and each of the loops is associated with guards $g$ (and $g'$)
that express the reachability of the loop head (and the end of the
loop body, respectively; see Section~\ref{sec:ssa}).
That is, suppose $f$ has $k$ loops
and $n_i$ denotes the number of lexicographic components for loop $i$,
then the termination argument
to prove termination of $f$ takes the form:
\[
\rrank^{\vec{n}}(\vx,\vxp) = \bigwedge_{i=1}^{k} g_i\wedge g'_i \Longrightarrow \lexrank^{n_i}_i(\vx,\vxp)
\]

While ranking techniques for mathematical integers use, e.g., Farkas' Lemma,
this is not applicable to bitvector operations. Thus, we use a synthesis
approach 
and extend it from
monolithic to lexicographic ranking functions.

We consider the class of lexicographic ranking functions generated by
the template where $\rank_i(\vx)$ is the product
$\vec{\rankparam}_i \vx$ with the row vector $\vec{\rankparam}_i$
of template parameters.
We~denote the resulting constraints for loop~$i$ as
$\lexranktempl_i^{n_i}(\vx,\vxp,\rankparamvec_i^{n_i})$ where
$\rankparamvec_i^{n_i}$ is the vector
$(\vec{\rankparam}_i^1,\dots,\vec{\rankparam}_i^{n_i})$.
The constraints for the ranking functions of a whole procedure are
$\rranktempl(\vx,\vxp,\vec{\rankparamvec}^{\vec{n}})$, where
$\vec{\rankparamvec}^{\vec{n}}$ is the vector
$(\rankparamvec_1^{n_1},\dots,\rankparamvec_k^{n_k})$.

Putting all this together, we obtain the following reduction of ranking
function synthesis to a first-order quantifier elimination problem over
templates:
$$
\exists \vec{\rankparamvec}^{\vec{n}}: \forall \vx,\vxp:
\inv(\vx)\wedge\trans(\vx,\vxp) \Longrightarrow \rranktempl(\vx,\vxp,\vec{\rankparamvec}^{\vec{n}})
$$

The parameters $\rankparamvec_i^{n_i}$ are vectors of vectors of bitvectors
extended by the special value $\top$ in order
to complete the lattice of ranking constraints
$\lexranktempl_i^{n_i}$.
We define $\lexranktempl_i^{n_i}(\vx,\vxp,\top) \eqdef \true$
indicating that no ranking function has been found for the given
template (``don't know'').
We write $\bot$ for the equivalence class of bottom elements for which
$\lexranktempl_i^{n_i}(\vx,\vxp,\rankparamvec_i^{n_i})$
evaluates to $\false$, meaning that the ranking function has not yet
been computed. For example, $\vec{0}$ is a bottom element.
Note that this intuitively corresponds to the meaning of $\bot$ and
$\top$ as known from invariant inference by abstract interpretation
(see Section~\ref{sec:inference}).

The algorithms for solving these formulae are described in
detail in~\cite{CDK+18}.
Moreover, \cite{CDK+18} explains algorithms for conditional
termination, i.e.\ programs that terminate only when certain
conditions on their inputs hold. The objective is to determine
these conditions.
Also, an approach to modular termination analysis using
procedure-level summaries is described in~\cite{CDK+18}.
2LS has an experimental implementation for these algorithms.
The works~\cite{CDK+18,Sch16} investigate how to solve approximate
solutions to these inter-procedural formulae that result
from modular analysis. Moreover, the works~\cite{CDK+18,MSS17} also look into
under-approximating backwards analysis in an inter-procedural
context.


\subsection{Non-Termination Analysis}\label{sec:nontermination}

2LS implements two techniques for \emph{proving
  non-termination}~\cite{MMSSVW18,Marticek17}.
Both of the approaches are relatively  simple, yet appear to be reasonably
efficient in many practical cases.

The first approach is based on finding \emph{singleton recurrent sets}.
All loops are unfolded $k$ times (with $k$ being incrementally
increased), followed by a check whether there is some loop $L$ and a program
configuration that can be reached at the head of $L$ after both $k'$ and $k$
unwindings for some $k' < k$.
Such a check can be easily formulated in 2LS as a formula over the SSA
representation of programs with loops unfolded $k$ times.
This technique is able to find lasso-shaped executions in which a loop
returns to the same program configuration every $k-k'$ iterations
after $k'$ initial iterations.

The second approach tries to reduce the number of unwindings by looking for
loops that generate an \emph{arithmetic progression} over every integer
variable.
More precisely, it looks for loops $L$ for which each integer variable
$x$ can be associated with a constant $c_x$ such that every iteration
of $L$ changes the value of $x$ to $x+c_x$, keeping non-integer
variables unchanged.
Two queries are used to detect such loops: the first one asks whether
there is a configuration $\overline{x}$ and a constant vector
$\overline{c}$ (with the vectors ranging over all integer variables
modified in the loop and constants from their associated bitvector
domains) such that one iteration of $L$ ends in the
configuration $\overline{x}+\overline{c}$, while the second makes sure
that there is no configuration $\overline{x}'$ over which one
iteration of $L$ would terminate in a~configuration other than
$\overline{x}'+\overline{c}$.
If such a loop $L$ and a constant vector $\overline{c}$ are found,
non-termination of $L$ can be proved as follows:
First, we gradually exclude each configuration $\overline{x}$
reachable at the head of $L$ for which there is some $k$ such that $L$
cannot be executed from $\overline{x}+k.\overline{c}$ (intuitively
meaning that $L$ cannot be executed $k+1$ times from $\overline{x}$).
Second, we check whether there remains some non-excluded configuration
reachable at the head of $L$.

The termination and non-termination analyses are run in parallel, and
the first definite answer is used.
Among the non-termination analyses, several rounds of unwinding
are first tried with the singleton recurrence set approach.
If that is not sufficient, the arithmetic progression approach is
tried.
If that does not succeed either, further rounds of unwinding with the
former approach are run.

\subsection{Memory Safety}\label{sec:memorysafety}

Analysing memory safety requires reasoning about values of pointers and about
the shape of the program heap. To this end, we introduce an abstract domain for
heap analysis---a so-called \emph{abstract shape domain}~\cite{MHSV18}. Memory
safety analysis is based on computing invariants in this domain and then
automatically generating assertions checking for absence of memory safety
errors, such as $\nullptr$ pointer dereference.

\subsubsection{Abstract Domain for Heap Analysis}\label{sec:shape_domain}

Similarly to all abstract domains in 2LS, the abstract shape domain has a form
of a \emph{template} describing the desired property. The shape of the heap is
defined by pointer links among memory objects and therefore our shape domain is
limited to the set $\ptr$ of all pointers defined by~\eqref{eq:pointers}.
More particularly, since the shape domain is used to infer loop invariants, we
limit it to the set $\ptr^{lb}$ of all \emph{loop-back} pointers. 

Let $L$ be the set of all loops in the analysed program. We define
\begin{equation}
  \ptr^{lb} = \ptr \times L.
\end{equation}
Elements of this set represent abstractions of values of individual pointers
returning from the end of loop bodies that are introduced in our program
representation (see Section~\ref{sec:ssa}).  We denote loop-back pointers $(p,
l) \in \ptr^{lb}$ by $\ssalb{p}{i}$ where $i$ is the program location of the
end of the loop $l$.

The shape domain over-approximates the \emph{may-point-to} relation between the
set $\ptr^{lb}$ and the set of all symbolic addresses $\addr$. We define the
form of the heap template to be the formula
\begin{equation}
\templ^S \equiv \bigwedge_{\ssaplb{i} \in \ptrlb} 
  \templ^S_{\ssaplb{i}}(d_{\ssaplb{i}}).
\end{equation}
The template is a conjunction of \emph{template rows}
$\templ^S_{\ssaplb{i}}$ where each row corresponds to a single loop-back
pointer $\ssaplb{i}$ and it describes the points-to relation of that pointer.
The parameter $d_{\ssaplb{i}} \subseteq \addr$ of the row (i.e., the 
\emph{abstract value of the row}) specifies the set of all addresses from the
set $\addr$ that $p$ may point to at the location $i$. The template row can be
therefore expressed as a disjunction of equalities between the loop-back
pointer and all possible addresses:
\begin{equation}
\templ^S_{\ssaplb{i}}(d_{\ssaplb{i}}) \equiv 
  (\bigvee_{a \in d_{\ssaplb{i}}} \ssaplb{i} = a)
\end{equation}

\noindent Computing an invariant in the given abstract domain allows 2LS to
characterize the shape of the program heap. For example, abstract values of
template rows corresponding to pointer fields of abstract dynamic objects
describe linked paths in the heap, such as linked segments. 

\subsubsection{Memory Safety Assertions}

In order to analyse memory safety, 2LS automatically instruments the analysed
program by assertions to check for typical memory safety
errors~\cite{MHSV18,MHSV19}. These are in particular pointer dereferencing
errors, \texttt{free} errors, and memory leaks.  We now describe the structure
of these assertions.

\subsubsection*{Dereferencing/Freeing a $\nullptr$ Pointer}

To check for this kind of errors, 2LS adds an assertion $p \neq \nullptr$ to
each program location where $*p$ or \texttt{free(p)} occurs. Since the shape
domain over-approximates the points-to relation, it is possible to soundly
prove absence of such errors. If an error is found, BMC can be used to check
whether it is spurious.

\subsubsection*{Dereferencing/Freeing a Freed Pointer}

We introduce a special variable $\freevar$ initialised to $\nullptr$ that is
used to track the possibly freed objects. Every call to \texttt{free(p)} in a
program location $i$ is replaced by a formula 
\begin{equation}
  \ssa{\freevar}{i} = \ssagfree{i} ? \ssa{p}{j} : \ssa{\freevar}{k}
\end{equation}
where $\ssa{p}{j}$ and $\ssa{\freevar}{k}$ are relevant versions of $p$ and
$\freevar$, respectively, valid at $i$, and $\ssagfree{i}$ is a free Boolean
variable. This formula represents a non-deterministic update of the value of
$\freevar$ by the freed address.

The shape domain is then used to over-approximate the set of all addresses that
$\freevar$ may point to, which is essentially the set of all possibly freed
memory objects. Proving \texttt{free} safety is then done by adding an
assertion $p \neq \freevar$ at each program location where $*p$ or
\texttt{free(p)} occurs.

The nature of the shape domain guarantees soundness of this approach, however,
using it for abstract dynamic objects is often very imprecise. This is because
freeing one of the concrete objects represented by the abstract one does not
mean that the rest of the represented objects cannot be safely dereferenced or
freed. This problem is resolved by modifying the representation of
\texttt{malloc} calls described in Section~\ref{sec:memops}.

In addition to the set $\dynobjset{i}$ of abstract dynamic objects used to represent
all objects allocated at $i$, we add one object $\dynobj{i}{co}$ to
$\dynobjset{i}$. The object can be non-deterministically chosen as the
malloc result (just like any other $\dynobj{i}{k}$), however, it is guaranteed
to represent a concrete object (i.e., it can be allocated only once). This is
achieved by an additional condition asserting that $\dynobj{i}{co}$ cannot be
allocated if there is a pointer pointing to it at the entry to the allocation
site $i$. Therefore, the \texttt{malloc} representation has the form
\begin{equation}
  p_i = (\ssagos{i, co} \wedge 
  \bigwedge_{p \in \ptr} p \neq \&\dynobj{i}{co} ) \,?\,
  \&\dynobj{i}{co} : 
  (\ssagosij{i}{1} \,?\, \&\dynobj{i}{1} : (\dots))
\end{equation}

\noindent Afterwards, it is only allowed to assign the address of the concrete
object $\dynobj{i}{co}$ to $\freevar$ at each allocation site $i$. Checks for
the free safety are also done on concrete objects only, which helps to avoid
the described imprecision. This approach remains sound since $\dynobj{i}{co}$
represents an arbitrary object allocated at $i$ and if safety can be proven
for it, it can be assumed to hold for all objects allocated at $i$.

\subsubsection*{Memory Leaks Safety}

Similarly to the previous section, the variable $\freevar$ is used to check for
safety from memory leaks. At the end of the program, we check whether there is
an object $\dynobj{i}{co}$ such that $\freevar \neq \&\dynobj{i}{co}$. If such
an object is found, a memory leak is present. However, proving absence from
memory leaks is only possible for loop-free programs (or for programs with all
loops fully unwound). This is because we do not track sequencing of abstract
objects representing concrete objects allocated at a single allocation site,
and our analysis typically sees that $\dynobj{i}{co}$ can be skipped in
deallocation loops, and hence remains inconclusive on the memory leaks.



\subsection{Domain Combinations}\label{sec:combination}

One of the main advantages of program verification implemented in 2LS is that
all abstract domains are required to have a common form of
\emph{templates}---quantifier-free first order formulae.  Thanks to this
feature, it is quite straightforward to create various compositions of
different domains while relying on the solver to do the heavy-lifting on the
domain operators combination and on mutual reduction of the domain abstract
values. In 2LS, we use two such combinations: \emph{product templates} and
\emph{power templates}, particularly their form called \emph{templates with
symbolic paths}~\cite{MHSV18}.

\subsubsection{Product Templates}

Product templates are one of the simplest forms of abstract domain combination
in 2LS. They are based on using a \emph{Cartesian product template} that
combines domains of various kinds side-by-side. This can be achieved by simply
taking a conjunction of their templates.

A very interesting combination of this form is the combination of the shape and
polyhedra abstract domains. It allows 2LS to analyse values of variables of
pointer and numerical type at the same time. This helps not only for analysing
program manipulating pointers and numbers at the same time but also opens a
possibility to reason about contents of data structures at the program heap.
Thanks to this combination, it is possible to verify programs such as the one
in Figure~\ref{fig:ex-heap-data}.


\subsubsection{Templates with Symbolic Paths}

Using simple templates of invariants, such as the described polyhedra templates
or the heap shape template, may not be precise enough to analyse some programs,
especially programs working with abstract dynamic objects. In such programs, it
is often required that an invariant computed for a loop $l$ distinguishes which
loops were or were not executed before reaching $l$. When working with abstract
dynamic objects allocated in loops, this allows one to distinguish situations
when an abstract dynamic object does not represent any really allocated object,
and therefore an invariant describing it is not valid.

With respect to this, in order to improve precision, 2LS introduces the
concept of \emph{symbolic loop paths}. A symbolic loop path expresses which
loops in the program were executed. Since 2LS uses loop-select guards to
capture the control flow through the loops, a symbolic path is simply a
conjunction of loop-select guard literals.

Formally, let $G^{ls}$ by the set of all loop-select guards in the analysed
program. A symbolic loop path $\pi$ is defined as
\begin{equation}
  \pi = \bigwedge_{g \in G^{ls}} l_g
\end{equation}
where $l_g$ is a literal of $g$, i.e., $g$ or $\neg g$. We denote the set of
all symbolic paths by $\Pi$. We also define a special path $\pi_\perp$
containing negative literals only. For this path, no loop invariant is computed
since no loops were executed for this path.

Having some template $\templ$ expressing an abstract domain (e.g. the shape
domain or even a product domain), we define the corresponding template with
symbolic paths as
\begin{equation}
  \templ^L \equiv \bigwedge_{\pi \in \Pi} \pi \implies \templ.
\end{equation}
This template can be viewed as a \emph{power template}---in the sense of power
domains---which assigns to each element of the base domain an element of the
exponent domain.

The algorithm for inference of an invariant $\inv^L$ from a symbolic path
template is shown in Algorithm~\ref{alg:sympaths}.
\begin{figure}[h!]
\begin{algorithmic}
  \STATE $\inv^L \gets \true$
  \FOR {$\pi \in \Pi$}
    \STATE add $\mathit{assert}(\pi)$ to solver
    \STATE $\inv_\pi \gets $ compute invariant from $\templ$
    \STATE remove $\text{assert}(\pi)$ from solver
    \IF {$(\pi \wedge \inv_\pi)$ is satisfiable}
      \STATE $\inv^L \gets \inv^L \wedge (\pi \implies \inv_\pi)$
    \ENDIF
  \ENDFOR
\end{algorithmic}
\caption{Algorithm for inferring an invariant from a symbolic paths template.}
\label{alg:sympaths}
\end{figure}

The algorithm computes a separate invariant $\inv_\pi$ in the inner domain
(using the template $\templ$) for each symbolic path $\pi$. To limit the
invariant computation for $\pi$ only, we assert that $\pi$ (a formula
expressing which loops are executed) holds during the computation. It may also
happen that, after computing $\inv_\pi$, the symbolic path is in fact not
reachable in the analysed program. Therefore, we check its reachability by
solving $\pi \wedge \inv_\pi$ in the context of the formula generated from the
SSA of the analysed program. If $\pi$ is reachable, then $\pi \implies
\inv_\pi$ is conjoined into the resulting invariant, otherwise $\inv_\pi$ is
discarded.






\section{Implementation and Architecture of 2LS}\label{sec:architecture}

2LS is based on the CPROVER framework.%
\footnote{\url{https://www.cprover.org}}
2LS is implemented in C++ and currently has around 25 KLOC
(not counting CPROVER itself).
2LS does not have any external dependencies beyond CPROVER
and can therefore be built on many platforms.
However, it is currently well-tested under Linux only.

The source code is available at \url{https://www.github.com/diffblue/2ls}
under the BSD license.
2LS was initially developed by Daniel Kroening and
Peter Schrammel in 2014 and has received contributions
by more than 15 contributors since then.

2LS is compatible with most of CPROVER tools such
as \texttt{goto-cc}, a drop-in replacement for GCC,
which can be used to build and link C source code using
the existing build system of the project to be analysed.
The resulting \textit{goto binary} can be passed to 2LS
and other CPROVER tools for analysis, e.g. \cbmc.

\subsubsection*{Architecture}

\begin{figure}[t]
\begin{tikzpicture}[line width=.5pt, >=latex]

\tikzstyle{boxnode} = [rectangle, draw,minimum height=3em, minimum width=7em,
text width=6.5em, align=center, node distance=10em]
\tikzstyle{domainnode} = [rectangle, draw, minimum height = 1.8em, minimum
  width = 8.8em, node distance = 2.5em, align=center]

\node (src) at (0,0) {source program};
\node [boxnode, right of = src] (goto) {GOTO generator};
\node [boxnode, right of = goto] (analysis) {Analysis + instrumentation};
\node [boxnode, right of = analysis] (ssa) {SSA generator};

\node [boxnode, below of = src] (unwind) {Unwinder};
\node [boxnode, below of = goto] (templ) {Template generator};
\node [boxnode, below of = analysis] (inv) {Invariant inference};

\node [domainnode, right = 2.1em of inv] (shape) {Shape domain};
\node [domainnode, above of = shape] (interval) {Interval domain};
\node [domainnode, below of = shape] (ranking) {Ranking domain};
\path (ranking) -- ($(ranking.south)+(0,-1.2)$) node [font=\Large, midway, sloped] {$\dots$};

\node [boxnode, below of = templ] (solver) {SAT solver};
\node [boxnode, below of = inv] (prop) {Property checker};
\node [boxnode, below = 2em of prop] (spurious) {Spuriousness checker};

\path [->] (src) edge (goto);
\path [->] (goto) edge (analysis);
\path [->] (analysis) edge (ssa);
\draw [->] (ssa) |- node [above, xshift = -10em] {SSA} ($(templ.north)+(0,1.3)$) -- (templ);
\path [->] (templ) edge (inv);
\path [->] (inv) edge (prop);
\path [->] (prop) edge (spurious);
\draw [->] (spurious) -| (unwind);
\path [->] (unwind) edge (templ);
\path [<->] (inv) edge (solver.north east);
\path [<->] (prop) edge (solver);
\path [<->] (spurious.north west) edge (solver.south east);
\path [<->] (interval.west) edge (inv.5);
\path [<->] (shape.west) edge (inv);
\path [<->] (ranking.west) edge (inv.-5);

\node[right = 4em of prop, yshift = 0.3em] (true) {
\begin{tikzpicture}
\fill[color=green!50!olive!80](0,.35) -- (.25,0) -- (1,.7) -- (.25,.15) -- cycle;
\end{tikzpicture}
};

\node[right = 3.8em of spurious, yshift = 1.6em] (false) {
\begin{tikzpicture}
\node[scale=2, red] {\ding{55}};
\end{tikzpicture}
};

\node[right = 4.5em of spurious, yshift = -1.6em, orange] (unknown) 
  {\LARGE \textbf ?};

\path [->] (prop.east) edge ($(prop.east)+(1.2,0)$);
\path [->] (spurious.east) edge ($(spurious.east)+(1.2,.35)$);
\path [->] (spurious.east) edge ($(spurious.east)+(1.2,-.35)$);

\node [right = 4em of ssa] (front) {Front-End};
\node [below of = front, node distance = 11em] (middle) {Middle-End};
\node [below of = middle, node distance = 12em] (back) {Back-End};

\end{tikzpicture}
\caption{2LS architecture overview.
}
\label{fig:architecture}
\end{figure}
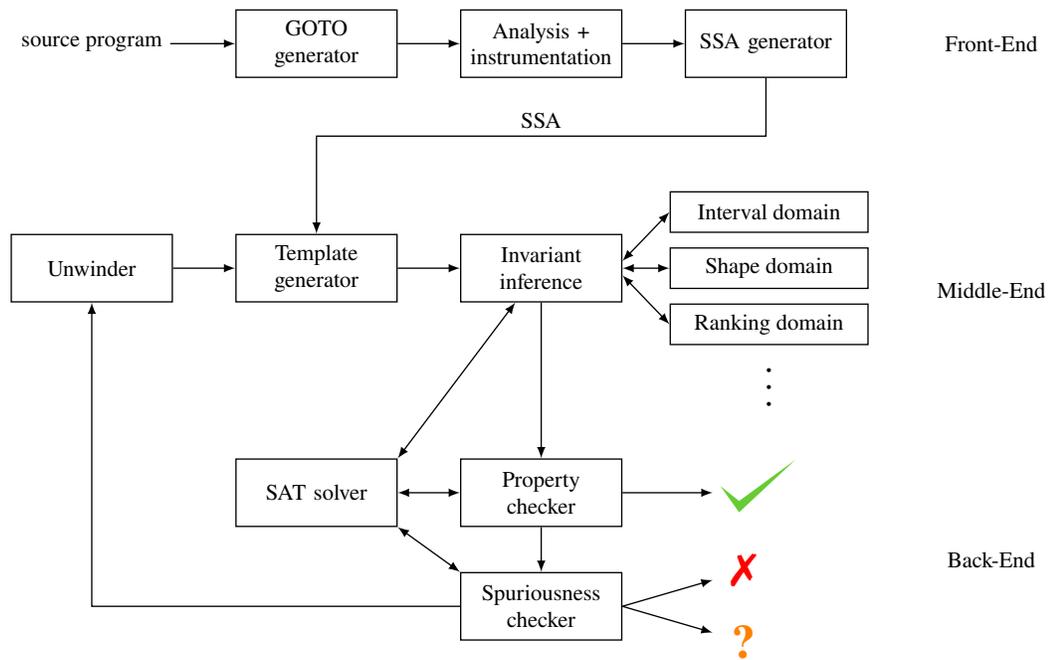

Similarly to other verification tools, 2LS follows the pipeline architecture
of compilers, consisting of a front-end for parsing and
type checking the source code; middle-end passes for
transforming and optimising the code based on an intermediate
representation (typically in the single static assignment form, SSA);
and a back-end, which, however, instead of code generation
performs the analysis and verification. 
An overview of the 2LS architecture is shown in Figure~\ref{fig:architecture}.

\section{Evaluations}\label{sec:evaluation}

We now present results of the 2LS framework in the International Competition on
Software Verification (SV-COMP). 2LS has competed in SV-COMP since 2016. In the
first part, we compare the score that 2LS achieved over the years in chosen
categories that require the kinds of analyses that 2LS implements. In the
second part, we present some alternative rankings of the participating tools
that demonstrate strengths of 2LS other than just the amount of programs
successfully verified (e.g., time or power consumption).

\subsection{Scores in SV-COMP}

In this section, we show how the score obtained by 2LS in SV-COMP evolved over
the years (from 2016 to 2022). We highlight the important milestones that
caused a significant rise in the score and we also mention some of the tool's
achievements.

Since SV-COMP consists of a large variety of verification tasks, we chose
several categories on which we demonstrate the tool's abilities.  We
concentrate on categories containing tasks whose verification requires program
analysis techniques that 2LS provides and that were described throughout this
chapter. In particular, we show the score in the following categories:
\begin{description}[font=\bfseries]
  \item[Loops + Floats] Here, we give the sum of the scores obtained in the
    \emph{Floats} and the \emph{Loops} sub-categories of the \emph{reachability
    safety} category%
    \footnote{In 2016, the SV-COMP categories layout was quite
    different from the current one, however, the \emph{Floats} and the
    \emph{Loops} categories already existed. \emph{Floats} was a standalone
    category, while \emph{Loops} was a sub-category of
    \emph{IntegersControlFlow}. The presented results from 2016 show score
    obtained in these (sub-)categories.}.
    Verification of programs in these categories requires inference of
    invariants over numerical variables, which 2LS does using the template
    polyhedra domain (Section~\ref{sec:tempoly}). In particular, we always
    used interval templates.

  \item[Heap] One of the strengths of 2LS is analysis of heap-manipulating
    programs. In SV-COMP, these are located in the \emph{ReachSafety-Heap}
    sub-category (where reachability of user-defined assertions is checked) and
    in the \emph{MemSafety} category (where tools verify absence of
    memory-safety errors). This part gives the sum of these two
    (sub-)categories. In order to analyse programs in this part, 2LS uses the
    abstract shape domain described in Section~\ref{sec:shape_domain}.

  \item[Termination] Besides verification of safety properties, 2LS allows to
    verify termination of programs by employing both termination
    (Section~\ref{sec:termination}) and non-termination
    (Section~\ref{sec:nontermination}) analysis and running them in parallel.
    In SV-COMP, termination is verified in the \emph{Termination} category.

  \item[Overall] The last kind of results that we show are those from the
    \emph{Overall} category that contains a (normalized) score obtained from
    verification of all task programs in SV-COMP.

\end{description}

\begin{figure}[h!]
\begin{tikzpicture}
  \pgfplotsset{set layers}
  \begin{axis} [
      width=10cm,
      height=7cm,
      /pgf/number format/1000 sep={},
      xmin=2016, xmax=2022,
      ymin=-1500, ymax=8000,
      xtick={2016,2017,2018,2019,2020,2021,2022},
      ytick={-1000,0,2000,4000,6000,8000},
      xlabel=Year,
      ylabel=Score,
      legend style={at={(0.03,0.96)},anchor=north west}
    ]

    \addplot coordinates {
      (2016, 272)(2017, 393)(2018, 390)(2019, 861)(2020, 785)(2021, 1016)(2022, 1153)
    };
    \label{int}
    \addlegendentry{Loops + Floats}

    \addplot coordinates {
      (2017, -1288)(2018, 277)(2019, 270)(2020, 439)(2021, 1239)(2022, 947)
    };
    \label{heap}
    \addlegendentry{Heap}

    \addplot coordinates {
      (2017, 624)(2018, 1522)(2019, 1279)(2020, 1264)(2021, 1315)(2022, 2178)
    };
    \label{term}
    \addlegendentry{Termination}

    \addplot coordinates {
      (2017, -1204)(2018, 4417)(2019, 4174)(2020, 4924)(2021, 6219)(2022, 7366)
    };
    \label{all}
    \addlegendentry{Overall}
  \end{axis}
\end{tikzpicture}
\caption{Chosen results in SV-COMP 2016-2022}
\label{fig:svcomp-results}
\end{figure}
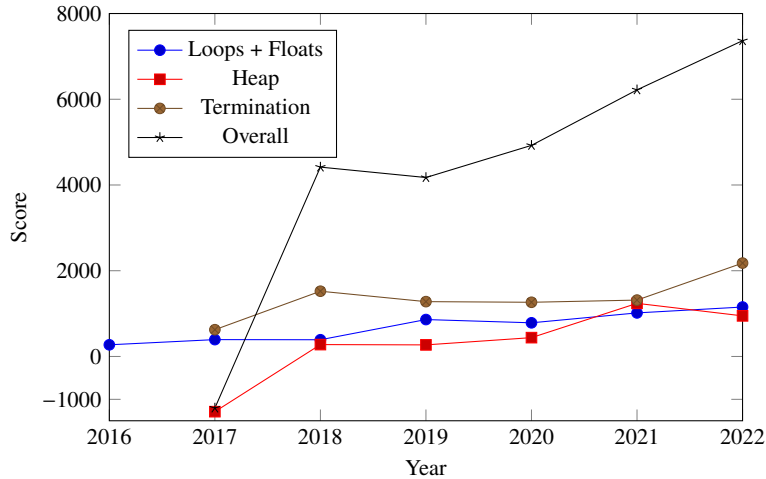

Figure~\ref{fig:svcomp-results} shows the evolution of the score that 2LS
achieved in these categories between years 2016 and 2022 (in 2016, 2LS did not
compete in most of the categories). The most remarkable improvement can be seen
in 2018, where the score in the heap and the termination categories improved
heavily, which caused also an improvement of the overall score. The reason for
this is that in 2018, the abstract shape domain and the non-termination
analysis were added to 2LS.

Next, we select some achievements of 2LS in the mentioned categories, which
support claims about the framework's strengths:
\begin{itemize}
  \item \textbf{first} place in \emph{Floats} and \textbf{second} place in
    \emph{Loops} in 2016,
  \item \textbf{third} place in \emph{Termination} in 2020, 2021, and 2022
  \item \textbf{fourth} place in \emph{Heap} in 2019.
\end{itemize}

Last, one of the strengths of 2LS is the capability to reason about the shape
and content of unbounded dynamic data structures at the same time (thanks to
abstract domain combinations presented in Section~\ref{sec:combination}). There
is no separate category for such programs; the only task programs requiring
such reasoning were added by our team in 2019~\cite{MHSV18}. In 2022, 2LS
remains one of the only two tools that are capable to verify majority of these
tasks.

\subsection{Alternative Rankings}

Besides the ability to soundly and correctly verify programs, there are other
properties that may support practical usefulness of a verification tool. In
this section, we compare 2LS to other verifiers using some alternative metrics,
in particular \emph{verification speed}, \emph{energy consumption}, and
\emph{correctness rate}.

\subsubsection{Speed of Verification in 2LS}

One property that can be observed from the SV-COMP results is that 2LS verifies
most of the tasks in a very short time, compared to other tools. We support
this claim by an experiment where we set a small time limit and then observe
how 2LS would compete against other tools in SV-COMP'22.
Table~\ref{fig:svcomp-speed} shows the position that 2LS would achieve in some
of the main categories if the time limit was set to 5 seconds.

\begin{table}[h!]
  \centering
  \renewcommand{\arraystretch}{1.2}
  \caption{Position of 2LS in some SV-COMP'20 categories with a 5s time limit.}
  \label{fig:svcomp-speed}
  \begin{tabular}{|c|c|c|c|c|c|}
    \hline
    ReachSafety & MemSafety & NoOverflows & Termination & Overall
    \\\hline
    5.          & 5.        & 1.          & 1.          & 2. \\\hline
  \end{tabular}
\end{table}

The table shows that 2LS would achieve a high position in all of the mentioned
categories. A notable result is the first place in the \emph{Termination} and
\emph{NoOverflows} categories and the second place in the \emph{Overall}
category.

\subsubsection{Power Consumption and Correctness Rate}

In recent years, the competition report of SV-COMP~\cite{svcomp20} provides two
\emph{alternative rankings} of verifiers that honor different aspects of the
verification process. These are in particular: 
\begin{itemize}
  \item \emph{Correct Verifiers} which ranks the verifiers by a so-called
    \emph{correctness rate}, which is a ratio of the number of incorrect results
    and the overall achieved score. 2LS finished third in this ranking in 2020
    with only 0.0016 errors per score point\footnote{In SV-COMP, a tool is
    granted 1 point for finding a bug in the program and 2 points for proving
    the program safe. The final score is then computed by normalisation of the
    scores obtained in the individual categories}.
  \item \emph{Green Verifiers} which ranks the verifiers by the amount of
    energy used to achieve a single score point. In 2020 and 2021, 2LS finished
    second in this ranking by using only 180\,J per score point.
\end{itemize}

\subsection{Evaluation Conclusion}

Overall, we summarize the observations obtained from the SV-COMP results. 2LS
is able to prove true properties as well as to find counterexamples to property
violations. Most of the tasks that it is capable to prove, it proves very fast,
mainly thanks to the incremental unwinding of a single SAT instance.

However, 2LS does not scale to more complex problems. Reasons for this are
multiple: templates may be too weak, invariant inference does not scale to
complex templates, or k-induction often does not succeed.

Moreover, 2LS still lacks support for verification of a number of features of
programs, such as array contents, concurrency, or recursion. These prevent 2LS
from better positions in some categories (MemSafety, Overall).

\section{Conclusions and Prospects}\label{sec:conclusion}

2LS is a verification system for C programs that is based on
a simple combination of bounded model checking, $k$-induction,
and template-based invariant inference.
The simplicity of the framework makes it easy to experiment
with new abstract domains.
2LS beats most other state-of-the-art tools in terms of
speed, but it still lagging behind in terms of robustness and
feature support.
For example, concurrency is not supported yet.
Also, recursion is not yet supported by its incremental
structural transformation algorithms.
Further improvements of the memory model will be required
to handle properties such as absence of memory leaks.
Invariant inference for arrays is another area of ongoing
work.


\end{document}